\documentclass[12pt]{article}
\usepackage{amsmath,amsfonts,amssymb,amsbsy,amsthm,mathtools}
\usepackage{hyperref}
\usepackage[margin=1in]{geometry}
\usepackage{graphicx}
\usepackage{subfigure}
\usepackage{multicol}
\usepackage{epsfig}
\usepackage{natbib}
\usepackage{bbm,bm,url,mathtools,color}
\graphicspath{{Figures/}} 

\newcommand{\vc}[1]{\bm{#1}}
\newcommand{\dm}{d}
\newcommand{\PP}{\mathbb{P}}
\newcommand{\diff}{\mathrm{d}}
\renewcommand{\Pr}{\mathbb{P}}

\newcommand{\abs}[1]{\left\lvert{#1}\right\rvert}
\newcommand{\1}{\mathbbm{1}}
\newcommand{\dto}{\xrightarrow{d}}
\newcommand{\pto}{\xrightarrow{\Pr}}

\begin{document}

\title{Nonparametric estimation of extremal dependence}

\author{Anna Kiriliouk, Johan Segers and Micha\l{} Warcho\l\\
  \small Universit\'e catholique de Louvain\\
  \small Institut de Statistique, Biostatistique et Sciences Actuarielles\\
  \small Voie du Roman Pays 20, B-1348 Louvain-la-Neuve, Belgium\\
  \small \texttt{\{anna.kiriliouk,johan.segers,michal.warchol\}@uclouvain.be}%
}
\maketitle
\begin{abstract}
There is an increasing interest to understand the dependence structure of a random vector not only in the center of its distribution but also in the tails. Extreme-value theory tackles the problem of modelling the joint tail of a multivariate distribution by modelling the marginal distributions and the dependence structure separately. For estimating dependence at high levels, the stable tail dependence function and the spectral measure are particularly convenient. These objects also lie at the basis of nonparametric techniques for modelling the dependence among extremes in the max-domain of attraction setting. In case of asymptotic independence, this setting is inadequate, and more refined tail dependence coefficients exist, serving, among others, to discriminate between asymptotic dependence and independence. Throughout, the methods are illustrated on financial data.
\end{abstract}

\section{Introduction}\label{introduction}
Consider a financial portfolio containing three stocks: JP Morgan, Citibank and IBM.
We download stock prices from \url{http://finance.yahoo.com} between January 1, 2000, and December 31, 2013, and convert them to weekly negative log-returns: if $P_1,\ldots,P_n$ is a series of stock prices, the negative log-returns are
\begin{equation*}
X_{i} = - \log{ \left( P_i / P_{i-1} \right) }, \qquad i = 1,\ldots,n.
\end{equation*}
These three series of negative log-returns will be denoted by the vectors $\vc{X}_i = (X_{i1}, X_{i2}, X_{i3})$ for $i=1, \ldots, 729$. By taking negative log-returns we force (extreme) losses to be in the upper tails of the distribution functions. It is such extreme values, and in particular their simultaneous occurrence, that are the focus of this chapter.

\begin{figure}[ht]
\centering
\subfigure{\includegraphics[width=0.32\textwidth]{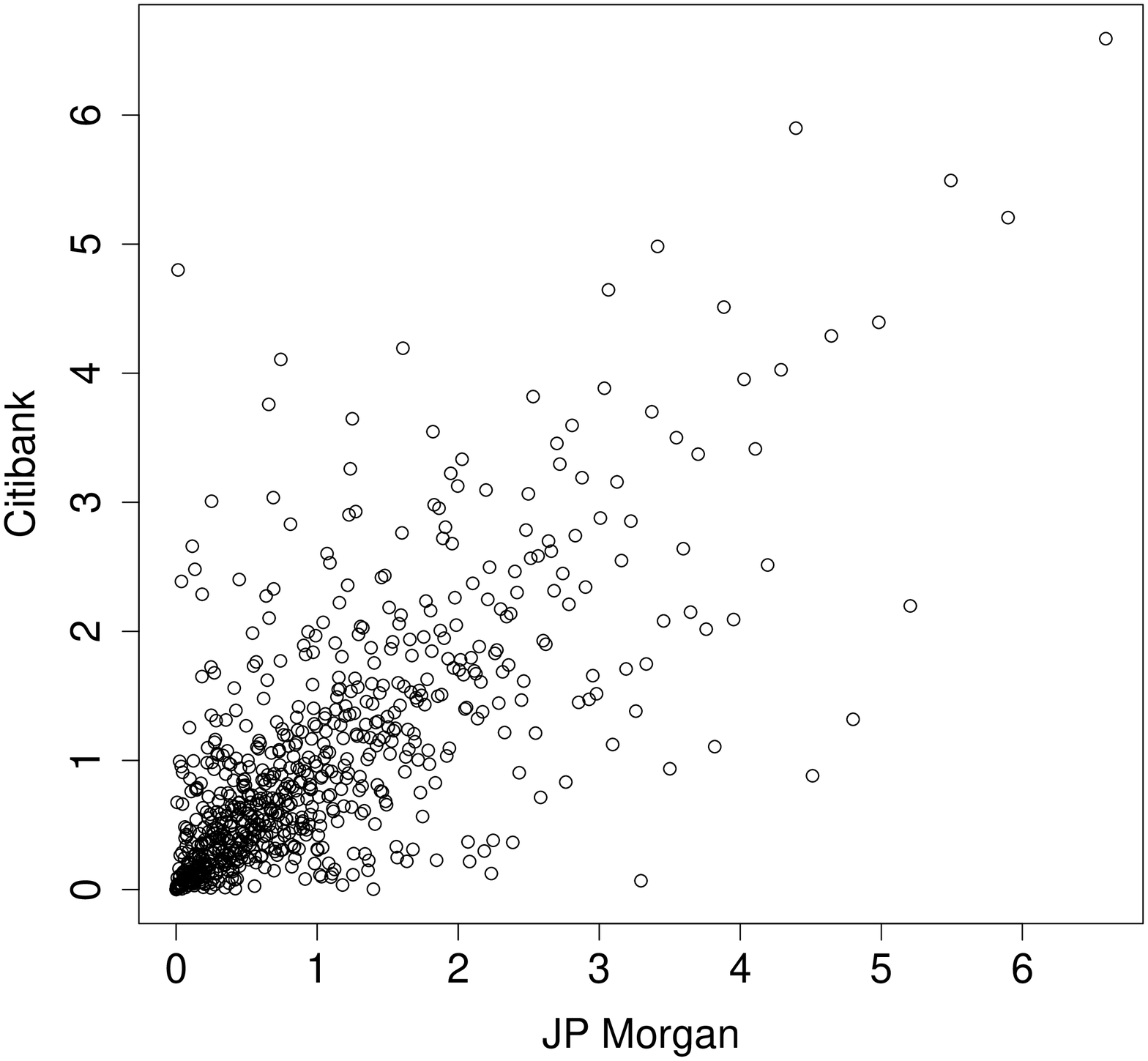}}
\subfigure{\includegraphics[width=0.32\textwidth]{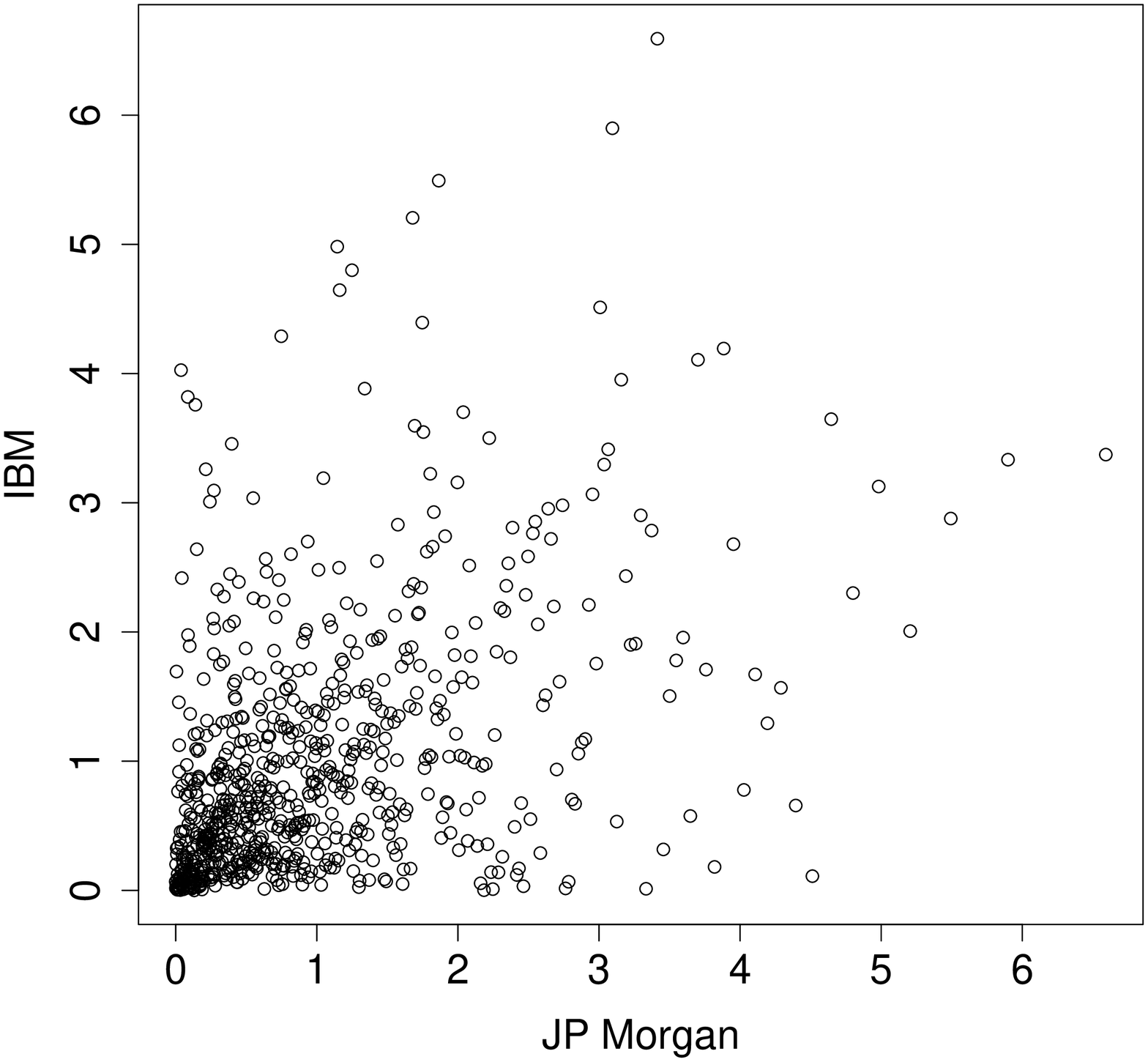}}  
\subfigure{\includegraphics[width=0.32\textwidth]{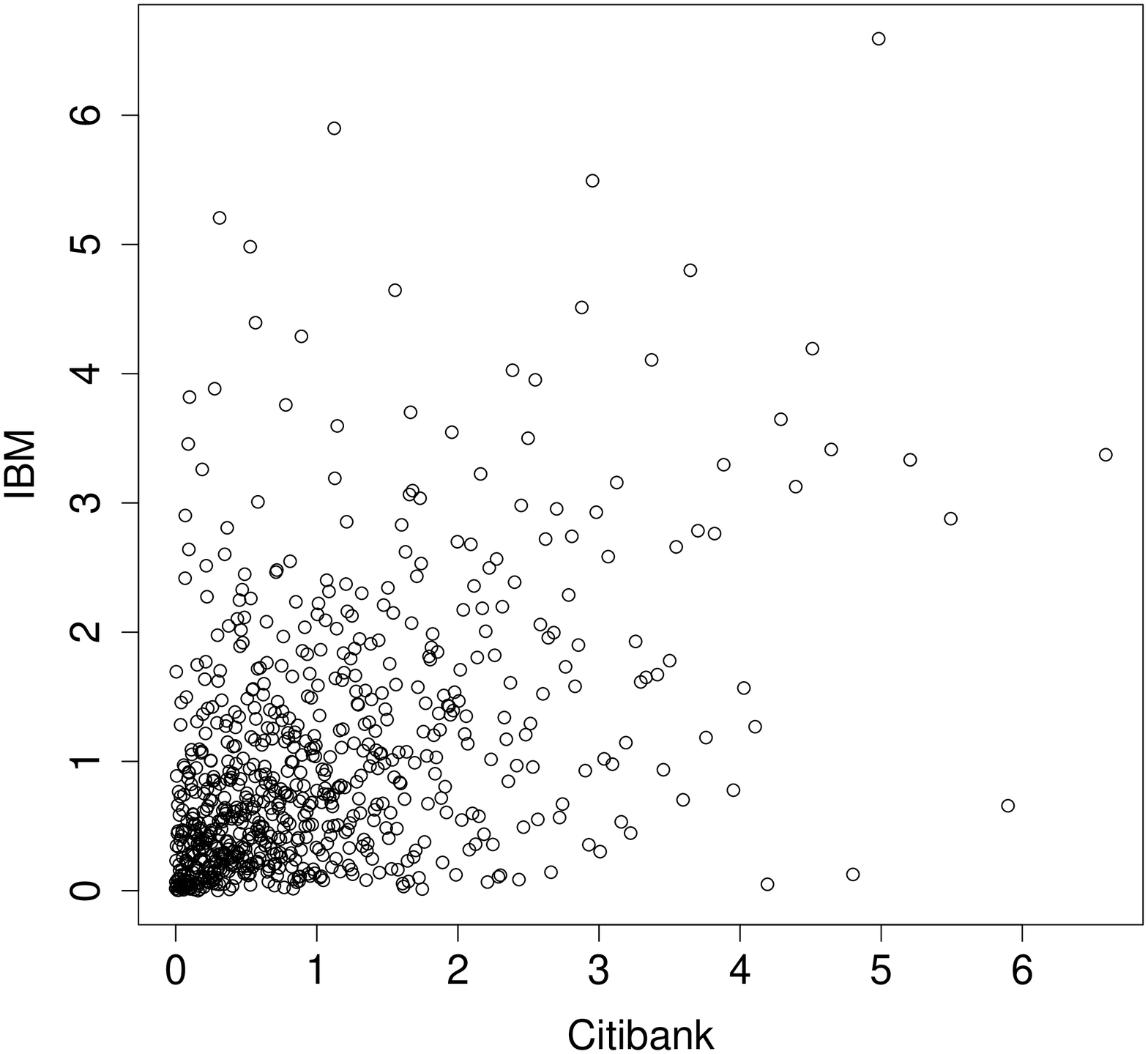}} 
\caption[Scatterplots of weekly stock price returns of JP Morgan, Citibank and IBM]
{Scatterplots of negative weekly log-returns of stock prices of JP Morgan, Citibank and IBM, plotted on the exponential scale.}
\label{fig:datadep}
\end{figure}

Figure~\ref{fig:datadep} shows the scatterplots of the three possible pairs of negative log-returns on the exponential scale. Specifically, we transformed the negative log-returns to unit-Pareto margins through
\begin{equation}
\label{eq:pareto}
  \widehat{X}_{ij}^* 
  \coloneqq \frac{1}{1 - \widehat{F}_j (X_{ij})}, 
  \qquad i \in \{1,\ldots,n\}, \; j \in \{1,\ldots,d\},
\end{equation}
(with $d = 3$) and plotted them on the logarithmic scale; the empirical distribution functions $\widehat{F}_j$ evaluated at the data are defined as
\begin{equation}
\label{eq:empiric}
  \widehat{F}_j (X_{ij}) 
  \coloneqq \frac{R_{ij,n} - 1}{n}, 
  \qquad i \in \{1,\ldots,n\}, \; j \in \{1,\ldots,d\},
\end{equation}
where $R_{ij,n}$ is the rank of $X_{ij}$ among $X_{1j},\ldots,X_{nj}$, i.e., $R_{ij,n} = \sum_{l=1}^n \1 \left\{ X_{lj} \leq X_{ij} \right\}$.

From Figure~\ref{fig:datadep}, we observe that joint occurrences of large losses are more frequent for JP Morgan and Citibank (left) than for JP Morgan and IBM (middle) or for Citibank and IBM (right). Given the fact that JP Morgan and Citibank are financial institutions while IBM is an IT company, this is no surprise. In statistical parlance, the pair JP Morgan versus Citibank exhibits the strongest degree of upper \emph{tail dependence}.

The purpose of this chapter is to describe a statistical framework for modelling such tail dependence. Specifically, we wish to model the tail dependence structure of a $d$-variate random variable $\vc{X} = (X_1, \ldots, X_d)$, with continuous marginal distribution functions $F_j(x_j) = \Pr(X_j \le x_j)$ and joint distribution function $F(\vc{x}) = \Pr(X_1 \le x_1, \ldots, X_d \le x_d)$. 

Preliminary marginal transformations of the variables 
or changes in the measurement unit should not affect the dependence model. Such invariance is obtained by the probability integral transform, producing random variables $F_1 (X_1), \ldots, F_d (X_d)$ that are uniformly distributed on the interval $(0, 1)$. Large values of $X_j$ correspond to $F_j(X_j)$ being close to unity, whatever the original scale in which $X_j$ was measured. In order to magnify large values, it is sometimes convenient to consider a further transformation to unit-Pareto margins via $X_j^* = 1 / \{ 1 - F_j(X_j) \}$; note that $\Pr[X_j^* > z] 
= 1/z$ for $z \ge 1$.

In practice, the marginal distributions are unknown and need to be estimated. A simple, robust way to do so is via the empirical distribution functions. The data are then reduced to the vectors of ranks as in \eqref{eq:pareto} and \eqref{eq:empiric}.

Starting point of the methodology is a mathematical description of the {upper tail dependence of the vector $\vc{X^*} = (X_1^*, \ldots, X_d^*)$ of standardized variables. There is a close connection with the classical theory on vectors of componentwise maxima, the multivariate version of the Fisher--Tippett--Gnedenko theorem \citep{fisher1928,gnedenko1943}. In the literature, there is a plethora of objects available for describing the dependence component of the limiting multivariate extreme-value distribution. Out of these, we have singled out the stable tail dependence function $\ell$ and the spectral measure $H$ (Section~\ref{s:prob}). The former is convenient as it yields a direct approximation of the joint tail of the distribution function of $\vc{X^*}$ and thus of $\vc{X}$; see \eqref{eq:tail} below. The latter has the advantage of representing tail dependence as the distribution of the relative magnitudes of the components of $\vc{X}^*$ given that the vector itself is `large'; see \eqref{eq:spectral:measure1} below.

Nonparametric inference on tail dependence can then be based on the stable tail dependence function and the spectral measure (Section~\ref{s:estim}). For convenience and in line with most of the literature, the description of the spectral measure estimators is limited to the bivariate case, although the generalization to the general, multivariate case is straightforward.

A weak point of the classical theory is its treatment of asymptotic independence, that is, when the tail dependence coefficient in \eqref{eq:chi} is equal to zero. Classically, this case is assimilated to the situation where the variables are exactly independent. Such a reduction applying to all bivariate normal distributions with pairwise correlation less than unity, it is clearly inadequate. More refined tail dependence coefficients as well as tests to distinguish asymptotic dependence from asymptotic independence are presented in Section~\ref{s:AI}.

The present chapter can serve as an introduction to nonparametric multivariate extreme value analysis. Readers looking for in-depth treatments of multivariate extreme-value distributions and their max-domains of attraction can consult, for instance, 
Chapter~5 in \cite{resnick1987} or Chapter~6 in \cite{dehaan2006}. Some other accessible introductions are the expository papers by \citet{dehaan1998} and \citet{segers2012}, Chapter~8 in \citet{beirlant2004}, and Part~II in \citet{falk:husler:reiss:2011}. The material on asymptotic independence is inspired on \citet{coles1999}.

\section{Tail dependence}
\label{s:prob}
\subsection{The stable tail dependence function}
\label{ss:prob:stdf}

Let $\vc{X} = (X_1,\ldots,X_d)$ be a random vector with continuous marginal distribution functions $F_j(x_j) = \Pr(X_j \le x_j)$ and joint distribution function $F(\vc{x}) = \Pr(X_1 \le x_1, \ldots, X_d \le x_d)$. To study tail dependence, we zoom in on the joint distribution of $(F_1(X_1), \ldots, F_d(X_d))$ in the neighbourhood of its upper endpoint $(1, \ldots, 1)$. That is, we look at
\begin{multline}
\label{eq:OR}
  1 - \Pr[ F_1(X_1) \le 1 - tx_1, \ldots, F_d(X_d) \le 1 - tx_d ] \\
  = \Pr[ F_1 (X_1) > 1 - t x_1 \text{ or } \ldots \text{ or }  F_d (X_d) > 1 - t x_d ],
\end{multline}
where $t > 0$ is small and where the numbers $x_1, \ldots, x_d \in [0, \infty)$ parametrize the relative distances to the upper endpoints of the $d$ variables. 
The above probability converges to zero as $t \to 0$ and is in fact proportional to $t$:
\begin{align*}
  \max(t x_1, \ldots, t x_d) & \leq \Pr[F_1 (X_1) > 1- t x_1 \text{ or } \ldots \text{ or } F_d (X_d) > 1 - t x_d] \\
  & \leq t x_1 + \cdots + t x_d.
\end{align*}
The \emph{stable tail dependence function}, $\ell : [0,\infty)^d \rightarrow [0,\infty)$, is then defined by
\begin{equation}
\label{eq:ell}
  \ell(\vc{x}) 
  \coloneqq \lim_{t \downarrow 0} t^{-1}
  \Pr[F_1(X_{1}) > 1 - tx_1 \text{ or } \ldots \text{ or } F_d (X_{d}) > 1 - t x_d],
\end{equation}
for $\vc{x} \in [0, \infty)^d$ \citep{huang1992, drees1998}. 
\index{stable tail dependence function}
The existence of the limit in \eqref{eq:ell} is an assumption that can be tested \citep{einmahl2006}.

The probability in \eqref{eq:OR} represents the situation where \emph{at least one} of the variables is large: for instance, the sea level exceeds a critical height at one or more locations. Alternatively, we might be interested in the situation where \emph{all} variables are large simultaneously. Think of the prices of all stocks in a financial portfolio going down together. The \emph{tail copula}, $R : [0, \infty)^d \to [0, \infty)$, is defined by
\begin{equation}
\label{eq:R}
  R(\vc{x}) 
  \coloneqq \lim_{t \downarrow 0} t^{-1} 
  \Pr[F_1(X_{1}) >  1 - tx_1, \ldots, F_d (X_{d}) > 1 - t x_d],
\end{equation}
for $\vc{x} \in [0,\infty]^d \setminus \{ (\infty,\ldots,\infty) \}$. 
\index{tail copula} 
Again, existence of the limit is an assumption. In the bivariate case ($d=2$), the functions $\ell$ and $R$ are directly related by $R(x,y) = x + y - \ell(x,y)$.
The difference between $\ell$ and $R$ is visualized in Figure~\ref{fig:ellRplot} for the log-returns of the stock prices of JP Morgan versus Citibank. From now on, we will focus on the function $\ell$ because of its direct link to the joint distribution function; see \eqref{eq:tail}.

\begin{figure}[ht]
\centering
\subfigure{\includegraphics[width=0.48\textwidth]{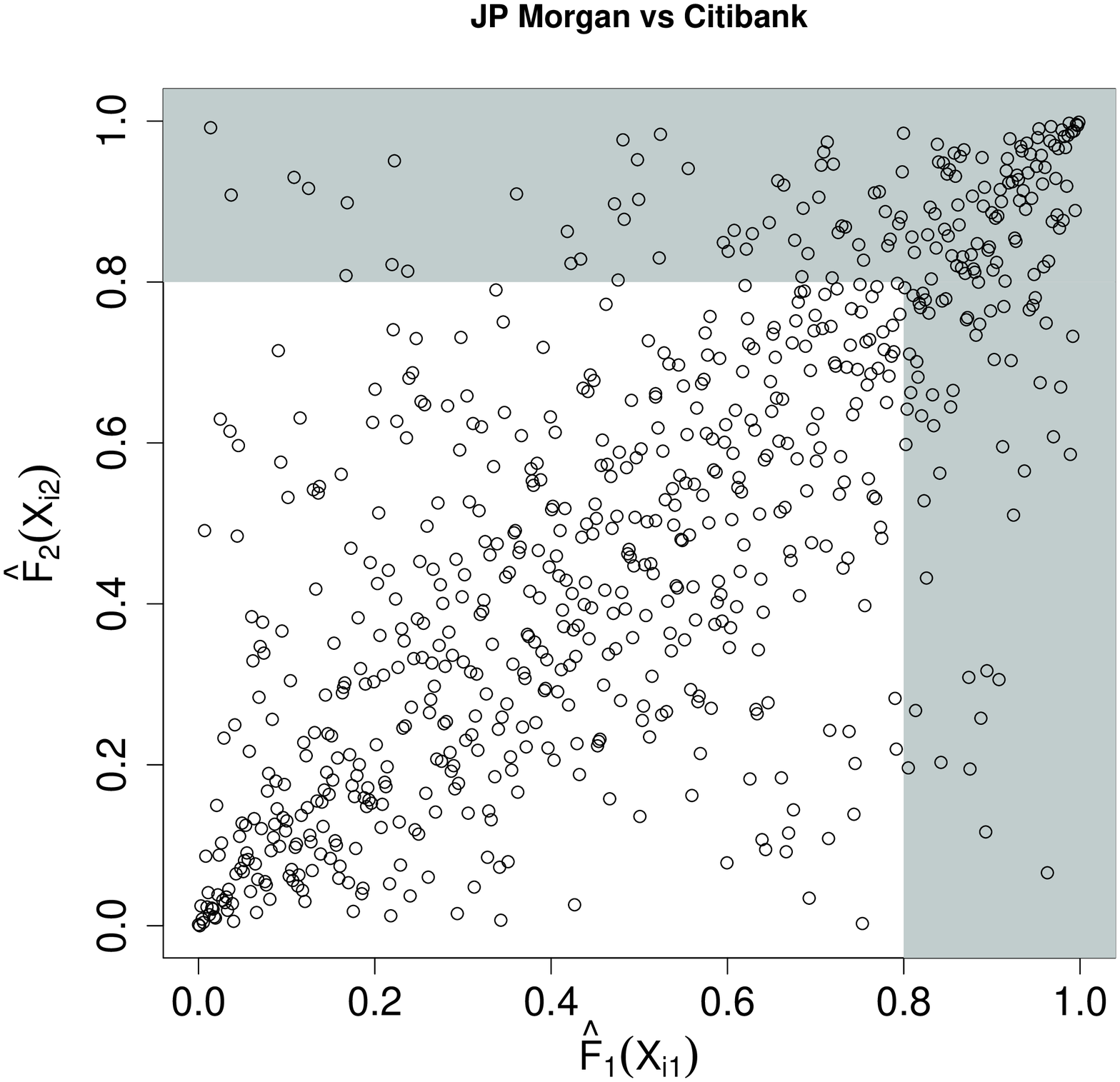}}
\subfigure{\includegraphics[width=0.48\textwidth]{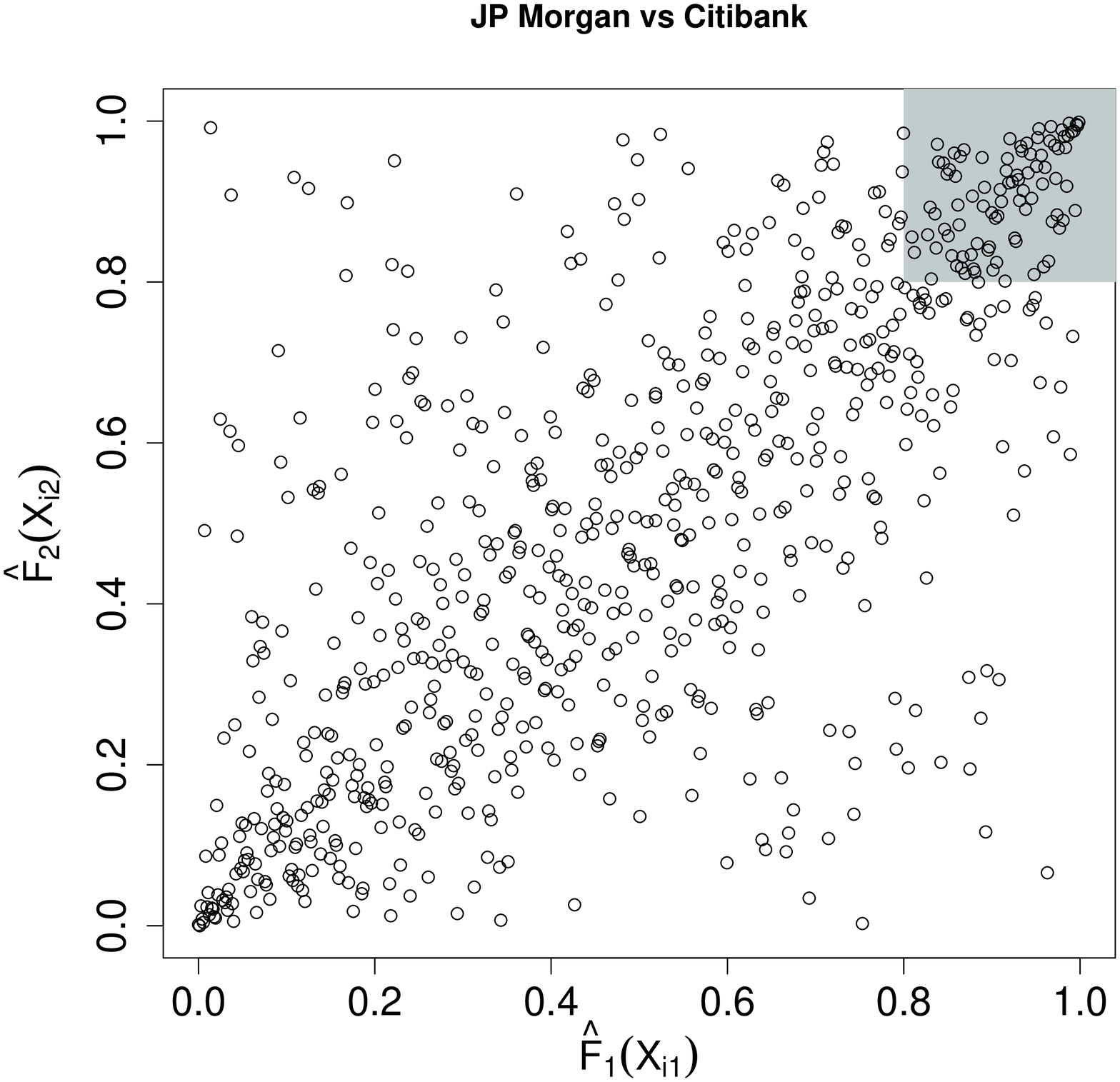}}    
\caption[The stable tail dependence function $\ell$ and the tail copula $R$]
{Scatterplots of negative weekly logreturns of stock prices of JP Morgan versus Citibank, transformed using ranks via \eqref{eq:empiric}. Left: the region where at least one variable is large, inspiring the definition of the stable tail dependence function $\ell$ in \eqref{eq:ell}. Right: the region where both variables are large, inspiring the definition of the tail copula $R$ in \eqref{eq:R}.}
\label{fig:ellRplot}
\end{figure}

It is convenient to transform the random variables $X_1, \ldots, X_j$ to have the unit-Pareto distribution via the transformations $X_j^* = 1 / \{1 - F_j(X_j)\}$ for $j = 1, \ldots, d$.
The existence of $\ell$ in \eqref{eq:ell} is equivalent to the statement that
the joint distribution function, $H_0$, of the random vector $\vc{X}^* = (X_1^*, \ldots, X_d^*)$ is in the max-domain of attraction of a $d$-variate extreme-value distribution, say $G_0$, with unit-Fr\'{e}chet margins, i.e., $G_{0j} (z_j) = \exp(-1/z_j)$ for $z_j > 0$. The link between $\ell$ and $G_0$ is given by
\begin{align*}
\ell(\vc{x}) 
& = \lim_{t \downarrow 0} t^{-1} 
  \left[ 
    1 - \Pr \left\{ X_1^* \le 1/(x_1 t), \ldots, X_d^* \le 1/(x_d t) \right\} 
  \right] \\
& = \lim_{n \to \infty} - \log{ H_0^n(n/x_1,\ldots,n/x_\dm) } \\
& =  -\log{ G_0 (1/x_1,\ldots,1/x_d)},
\end{align*}
so that $G_0 (\vc{z}) = \exp{\{-\ell(1/z_1,\ldots,1/z_d)\}}$.
Since, for ``large'' $z_1,\ldots,z_d$ we have
\begin{equation}
\label{eq:tail}
H_0(\vc{z}) 
\approx 
\exp \left\{ 
  - \frac{1}{n} \, \ell \left(  \frac{n}{x_1} , \ldots, \frac{n}{x_d} \right) 
\right\}
=
\exp \{ - \ell (1/z_1, \ldots, 1/z_d) \},
\end{equation}
we see that estimating the stable tail dependence function $\ell$ is key to estimation of the tail of $H_0$, and thus, after marginal transformation, of $F$. If, in addition, the marginal distributions $F_1,\ldots,F_d$ are in the max-domains of attraction of the univariate extreme-value distributions $G_1,\ldots,G_d$, then $F$ is in the max-domain of attraction of the $d$-variate extreme-value distribution $G(\vc{x}) = \exp\{ - \ell( - \log G_1(x_1), \ldots, - \log G_d(x_d) ) \}$.

Every stable tail dependence function $\ell$ has the following properties:
\begin{enumerate}
\item 
$\max{(x_1,\ldots,x_d)} \leq \ell(\vc{x}) \leq x_1 + \cdots + x_{d}$, and in particular $\ell(0,\ldots,0,x_j,0,\ldots,0) = x_j$ for $j=1,\ldots,d$;
\item 
convexity, that is, $\ell \bigl(t \vc{x} + (1-t) \vc{y} \bigr) \leq t \, \ell(\vc{x}) + (1-t) \, \ell(\vc{y})$, for $t \in [0,1]$;
\item 
order-one homogeneity: $\ell(ax_1,\ldots, ax_{\dm}) = a \, \ell(x_1,\ldots,x_{\dm})$, for $a > 0$;
\end{enumerate}
\citep[page 257]{beirlant2004}. When $d = 2$, these three properties characterize the class of stable tail dependence functions. When $d \ge 3$, a function satisfying these properties is not necessarily a stable tail dependence function. 

The two boundary cases are complete dependence, $\ell( \vc{x} ) = \max(x_1, \ldots, x_d)$, and independence, $\ell(\vc{x}) = x_1 + \cdots + x_d$. Another example is the \emph{logistic model},
\begin{equation}
\label{eq:logistic}
\ell (\vc{x} ; \theta) 
= \left( x_1^{\theta} + \cdots + x_d^{\theta} \right)^{1/ \theta}, 
\qquad \theta \in [1,\infty).
\end{equation}
As $\theta \rightarrow 1$, extremes become independent and as $\theta \rightarrow \infty$, extremes become completely dependent. The copula of the 
extreme-value distribution $G_0$ with this 
stable tail dependence function is the Gumbel--Hougaard copula.

The stable tail dependence function is related to two other functions:
\begin{itemize}
\item
The \emph{exponent measure function} $V(\vc{z}) \coloneqq - \log G_0(\vc{z}) = \ell(1/z_1, \ldots, 1/z_d)$, for $\vc{z} \in (0, \infty]^d$ \citep{coles1991}.
\index{exponent measure}
\item
The \emph{Pickands dependence function} $A$, which is just the restriction of $\ell$ to the unit simplex $\{ \vc{x} \in [0, \infty)^d : x_1 + \cdots + x_d = 1 \}$ \citep{pickands1981}. In dimension $d = 2$, 
it is common to set $A(t) = \ell(1 - t, t)$ for $t \in [0, 1]$.
\index{Pickands dependence function}
\end{itemize}

\subsection{The angular or spectral measure}
\label{ss:prob:H}

Another insightful approach for modelling extremal dependence of a $\dm$-dimensional random vector is to look at the contribution of each of the $\dm$ components to their sum conditionally on the sum being large. 

Consider first the bivariate case. In Figure~\ref{fig:spec:measure1} we show pseudo-random samples from the Gumbel copula, which is related to the logistic model~\eqref{eq:logistic}, at two different values of the dependence parameter $\theta$: 
strong dependence ($\theta = 2.5$) at the top and weak dependence ($\theta = 1.25$) at the bottom.
For convenience, the axes in the scatter plots on the left are on the logarithmic scale. The points for which the sum of the coordinates belong to the top-10\% are shown in black. 
In the middle panels we plot the ratio of the first component to the sum (horizontal axis) against the sum itself (vertical axis). On the right, finally, histograms are given of these ratios for those points for which the sum exceeded the threshold in the plots on the left and in the middle. Strong extremal dependence leads to the ratios being close to $0.5$ (top), whereas low extremal dependence leads to the ratios being close to either $0$ or $1$ (bottom).

We learn that the distribution of the ratios of the components to their sum, given that the sum is large, carries information about dependence between large values. This distribution is called the \emph{angular} or \emph{spectral measure} \citep{dehaan:resnick:1977} and is denoted here by $H$. 
\index{spectral measure}
\index{angular measure|see {spectral measure}}
The reference lines in Figure~\ref{fig:spec:measure1} represent the spectral density associated with the Gumbel copula with parameter $\theta>1$, that is,
\[
  \frac{\diff H_\theta}{\diff w}(w)
  = 
  \frac{\theta-1}{2} \, 
  \{w(1-w)\}^{-1-\theta} \, 
  \bigl\{ w^{-\theta}+(1-w)^{-\theta} \big\}^{1/\theta-2},
  \qquad 0 < w < 1.
\] 

In Figure~\ref{fig:spec:measure2} we show the same plots as in Figure~\ref{fig:spec:measure1} but now for the pairs of negative log-returns of the stock prices of JP Morgan versus Citibank (top) and JP Morgan versus IBM (bottom). For each of the two pairs we do the following. Let $(X_i, Y_i)$, $i = 1, \ldots, n$, denote a bivariate sample. First, we transform the data to the unit-Pareto scale using and \eqref{eq:pareto} and \eqref{eq:empiric}, i.e., 
\begin{equation}
\label{eq:transPareto}
  \widehat{X}_i^* = n / ( n + 1 - R_{i,X} )
  \quad \text{and} \quad
  \widehat{Y}_i^* = n / ( n + 1 - R_{i,Y} ), 
  \qquad i=1,\ldots,n.
\end{equation}
Here $R_{i,X}$ is the rank of $X_{i}$ among $X_{1},\ldots,X_{n}$ and $R_{i,Y}$ is the rank of $Y_{i}$ among $Y_{1},\ldots,Y_{n}$. Next, we construct
an approximate sample from the spectral measure $H$
by setting 
\begin{equation}
\label{eq:transPolar}
 \widehat{S}_i = \widehat{X}_i^* + \widehat{Y}_i^* \quad \text{ and } \quad
  \widehat{W}_i= \widehat{X}_i^* / ( \widehat{X}_i^* + \widehat{Y}_i^* ), 
 \qquad i=1,\ldots,n.
\end{equation}
The data in the pseudo-polar representation \eqref{eq:transPolar} are shown in the middle plots. The histograms of the ratios $\widehat{W}_i$ for those points for which the sum $\widehat{S}_i$ is above the $95\%$ sample quantile are displayed on the right-hand side. The solid black lines show the smoothed spectral density estimator in~\eqref{h.smooth} below. 

For JP Morgan versus Citibank, the ratios are spread equally across the unit interval, suggesting extremal dependence. In contrast, for JP Morgan versus IBM, the spectral density has peaks at the boundaries $0$ and $1$, suggesting weak extremal dependence.

\begin{figure}[ht]
\centering
\subfigure{\includegraphics[width=0.32\textwidth]{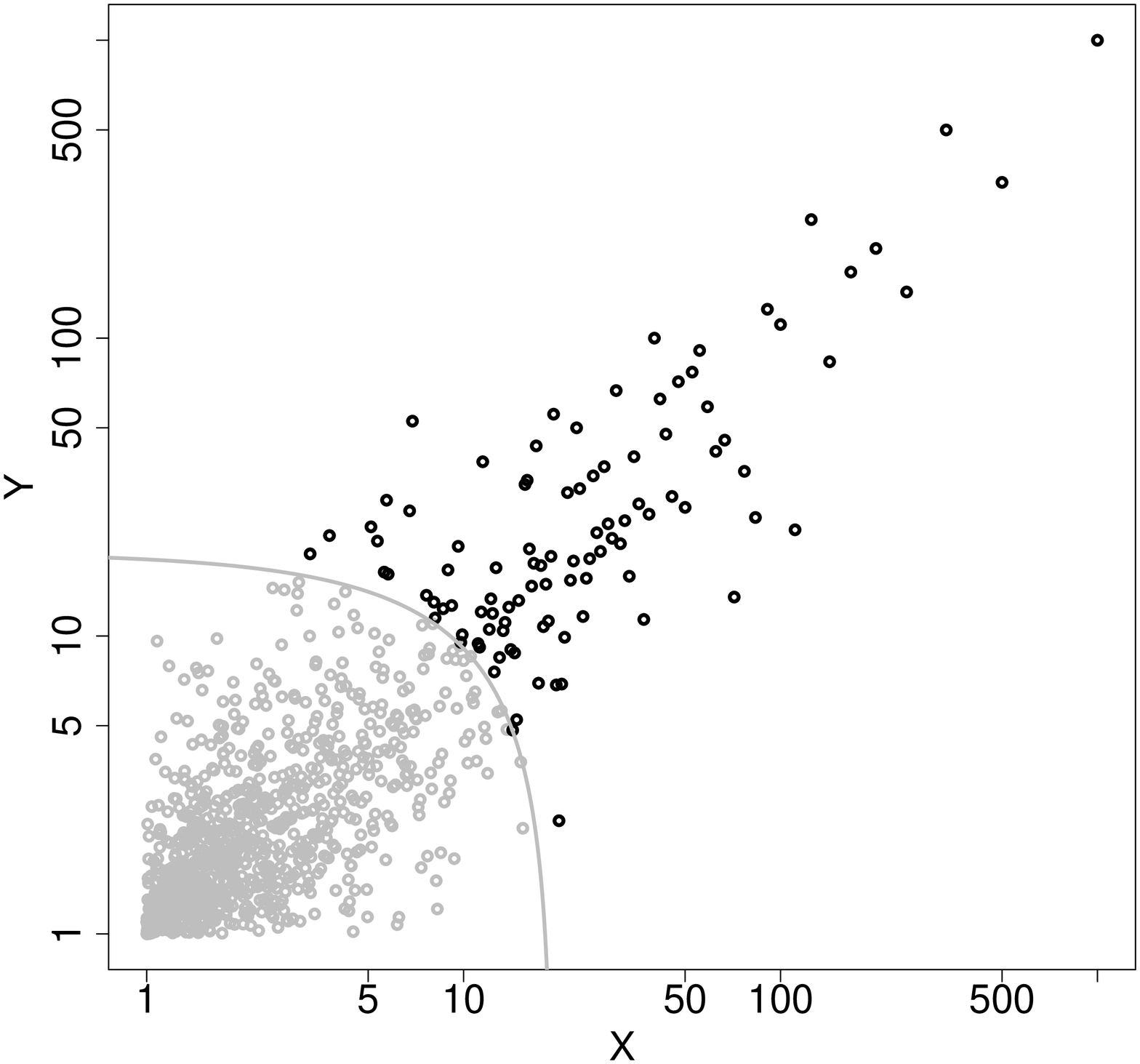}}
\subfigure{\includegraphics[width=0.32\textwidth]{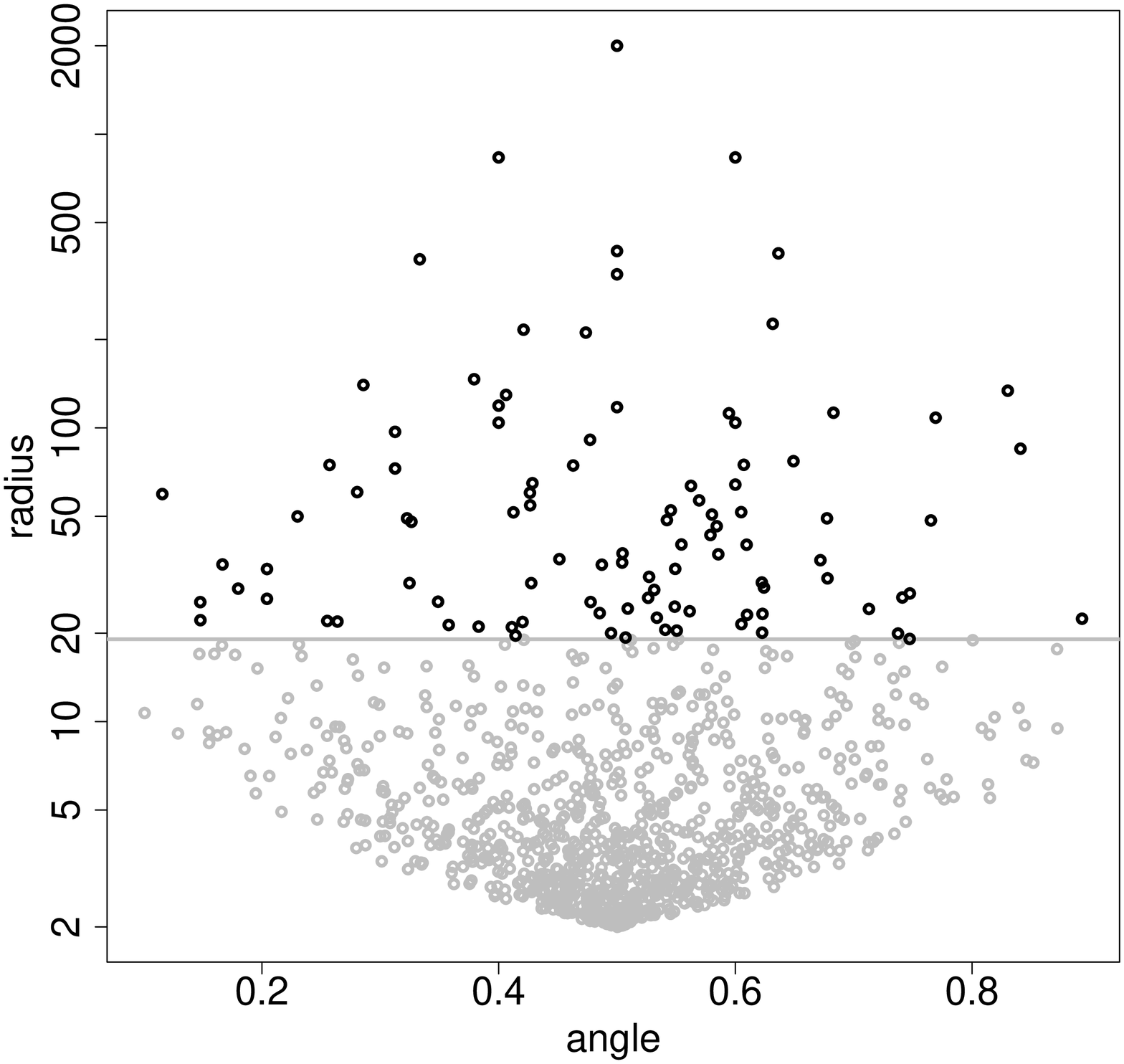}}  
\subfigure{\includegraphics[width=0.32\textwidth]{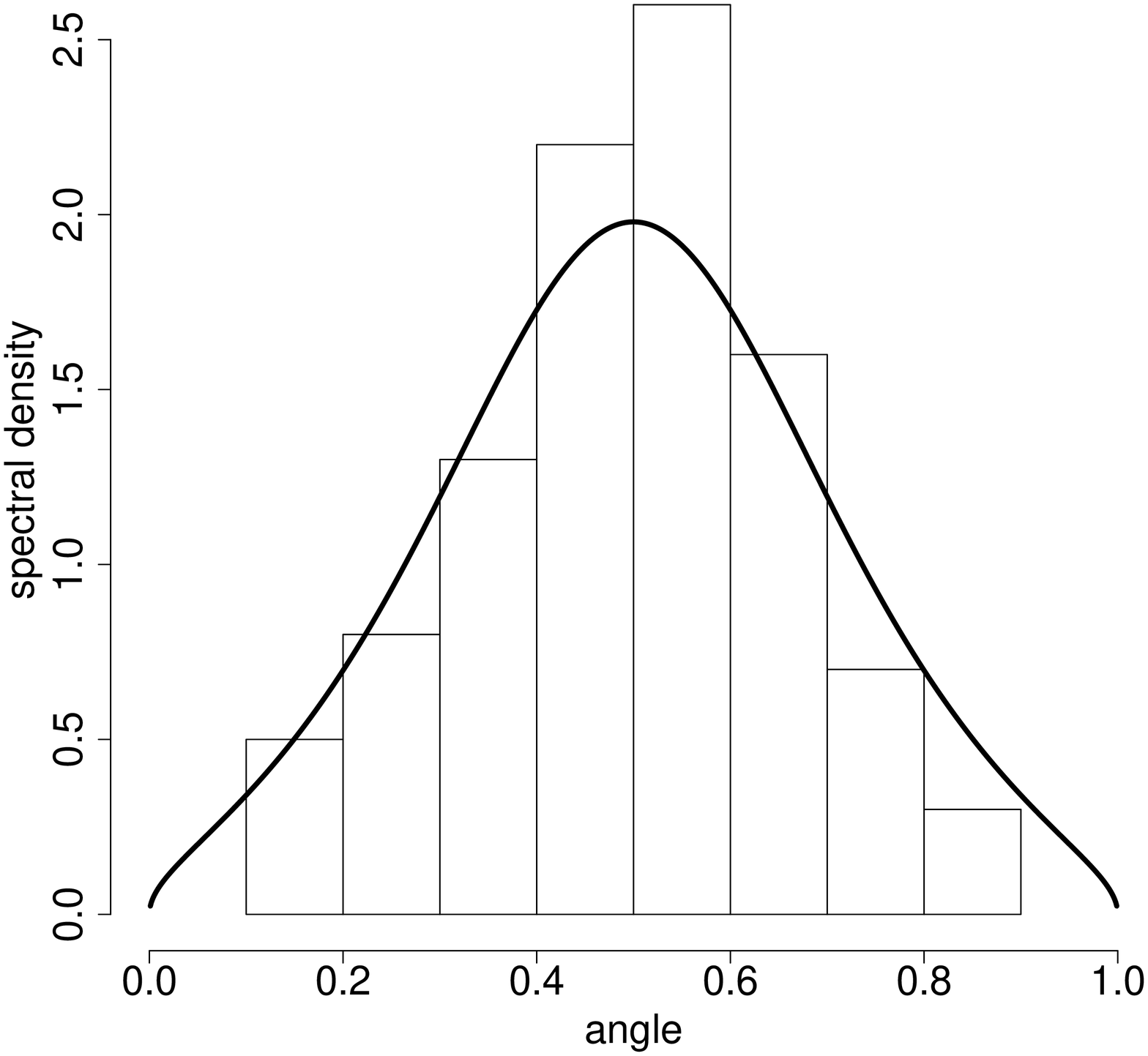}} \\
\subfigure{\includegraphics[width=0.32\textwidth]{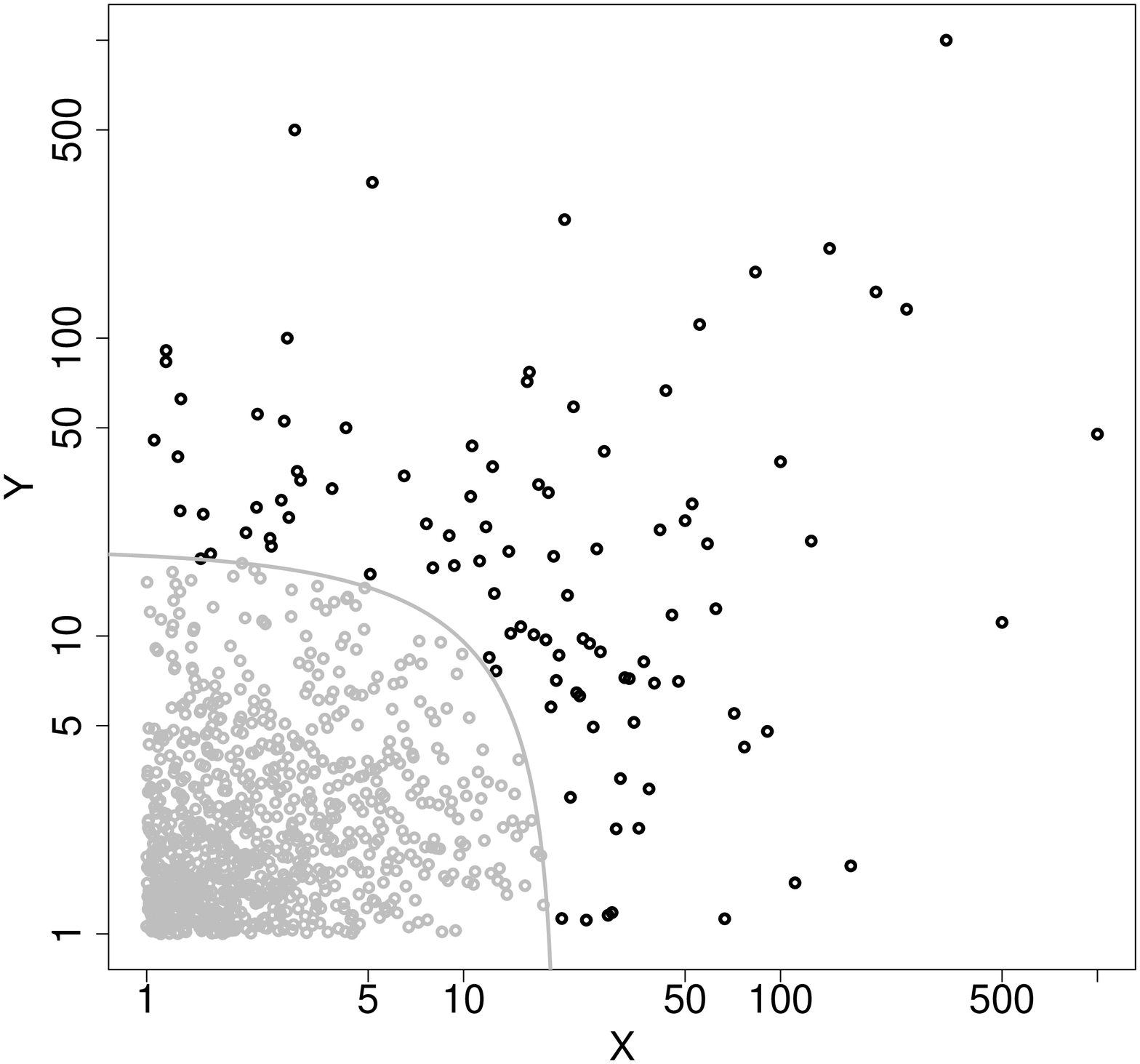}}
\subfigure{\includegraphics[width=0.32\textwidth]{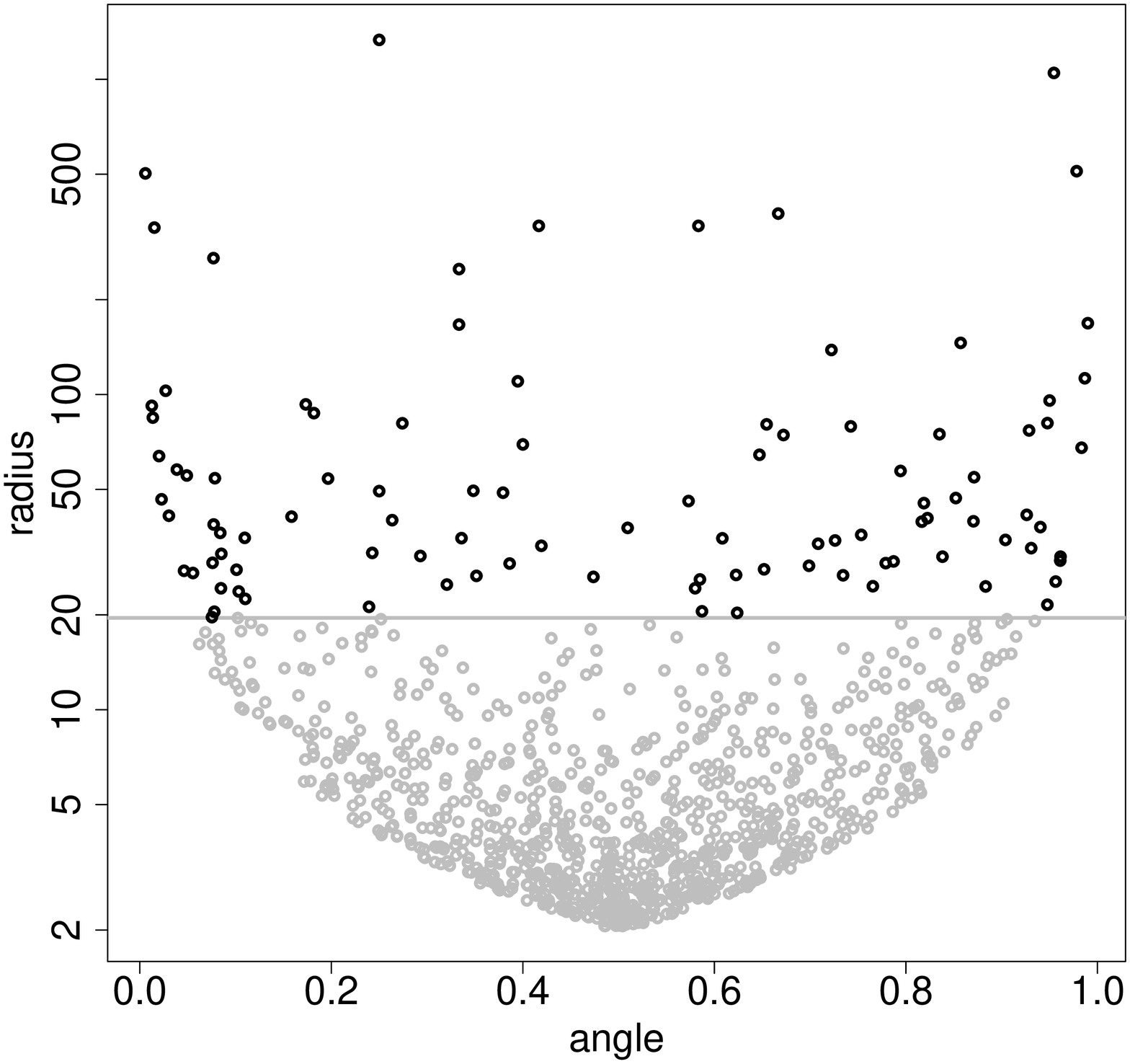}}  
\subfigure{\includegraphics[width=0.32\textwidth]{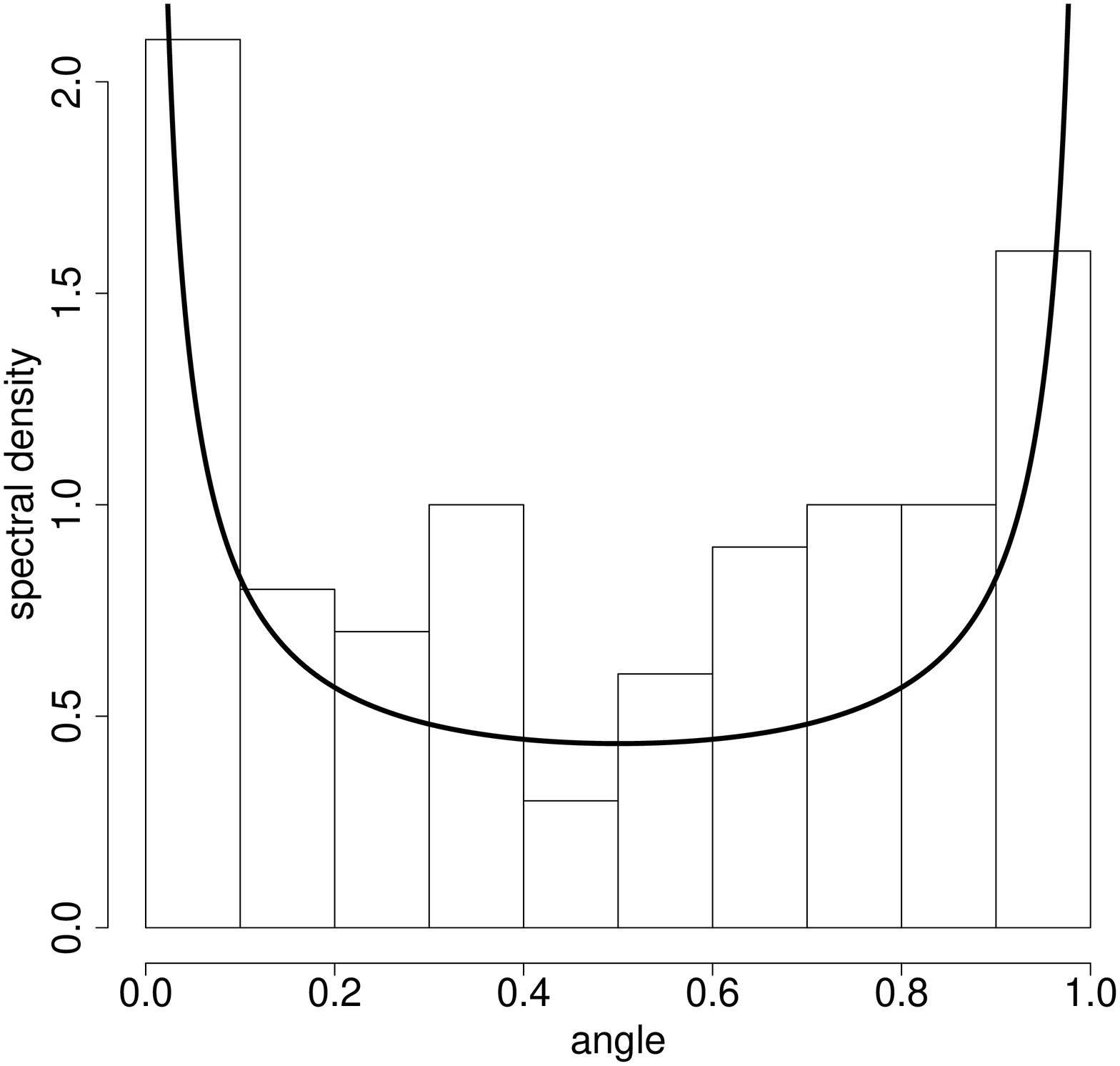}} 
\caption[Transformation into pseudo-polar coordinates for a random sample from the Gumbel copula]{Pseudo-random samples from the Gumbel copula; $\theta=2.5$ (top) and $\theta=1.25$ (bottom). Left: the data transformed to unit-Pareto margins (log scale); middle: further transformation into pseudo-polar coordinates; right: histogram of the angles of the extreme pairs and the true spectral density (solid line).}
\label{fig:spec:measure1}
\end{figure}

\begin{figure}[ht]
\centering
\subfigure{\includegraphics[width=0.32\textwidth]{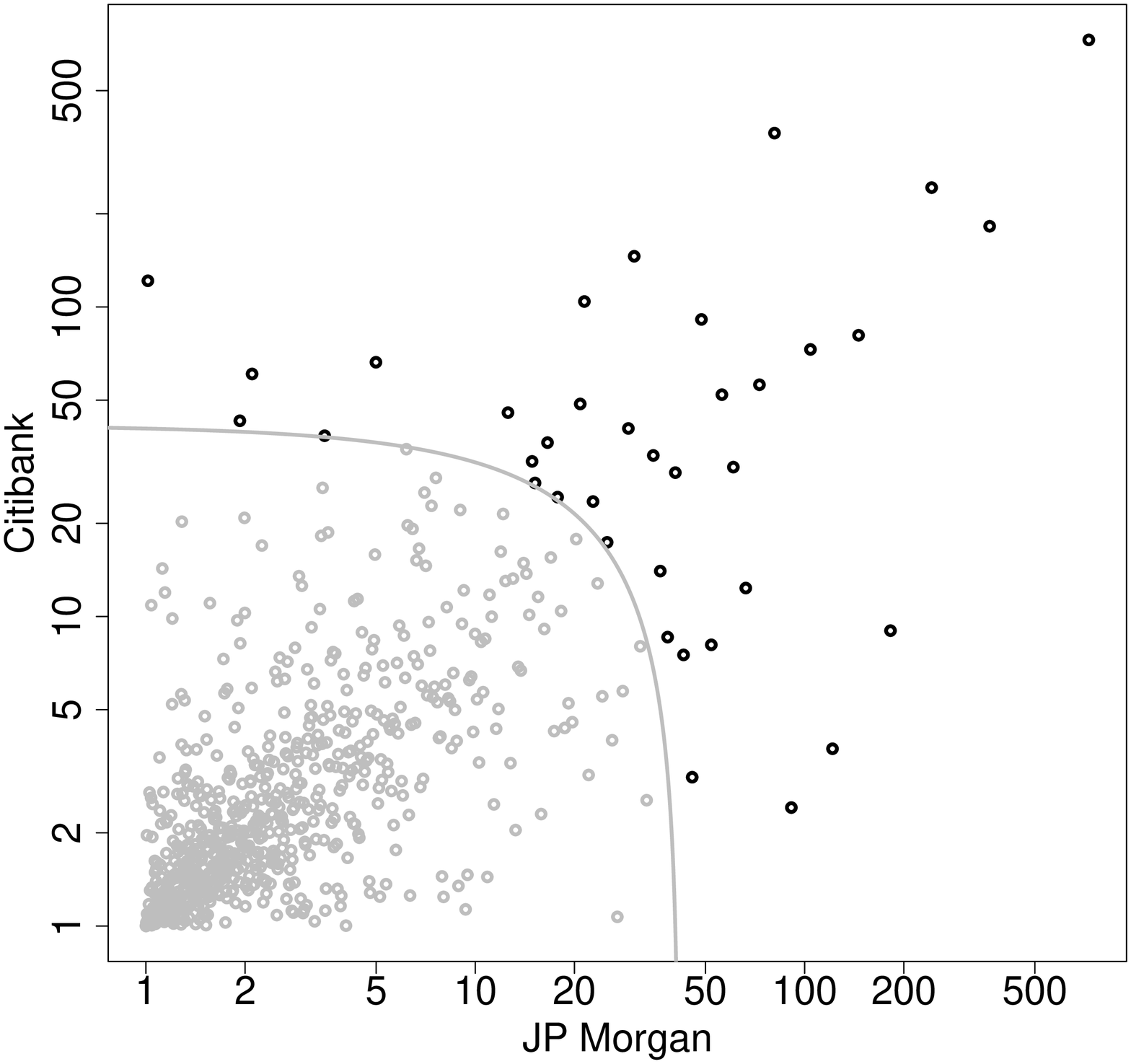}}
\subfigure{\includegraphics[width=0.32\textwidth]{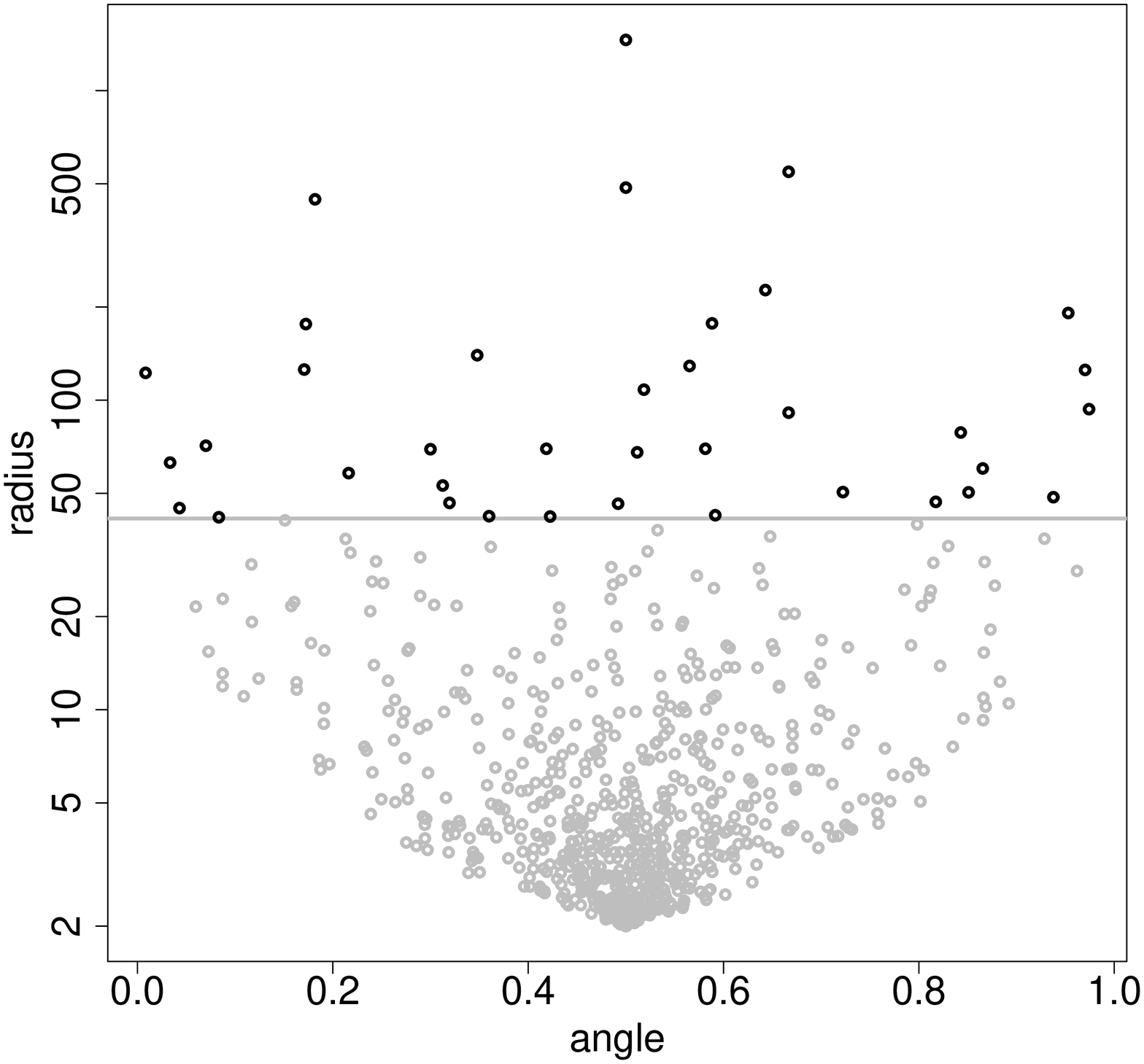}}  
\subfigure{\includegraphics[width=0.32\textwidth]{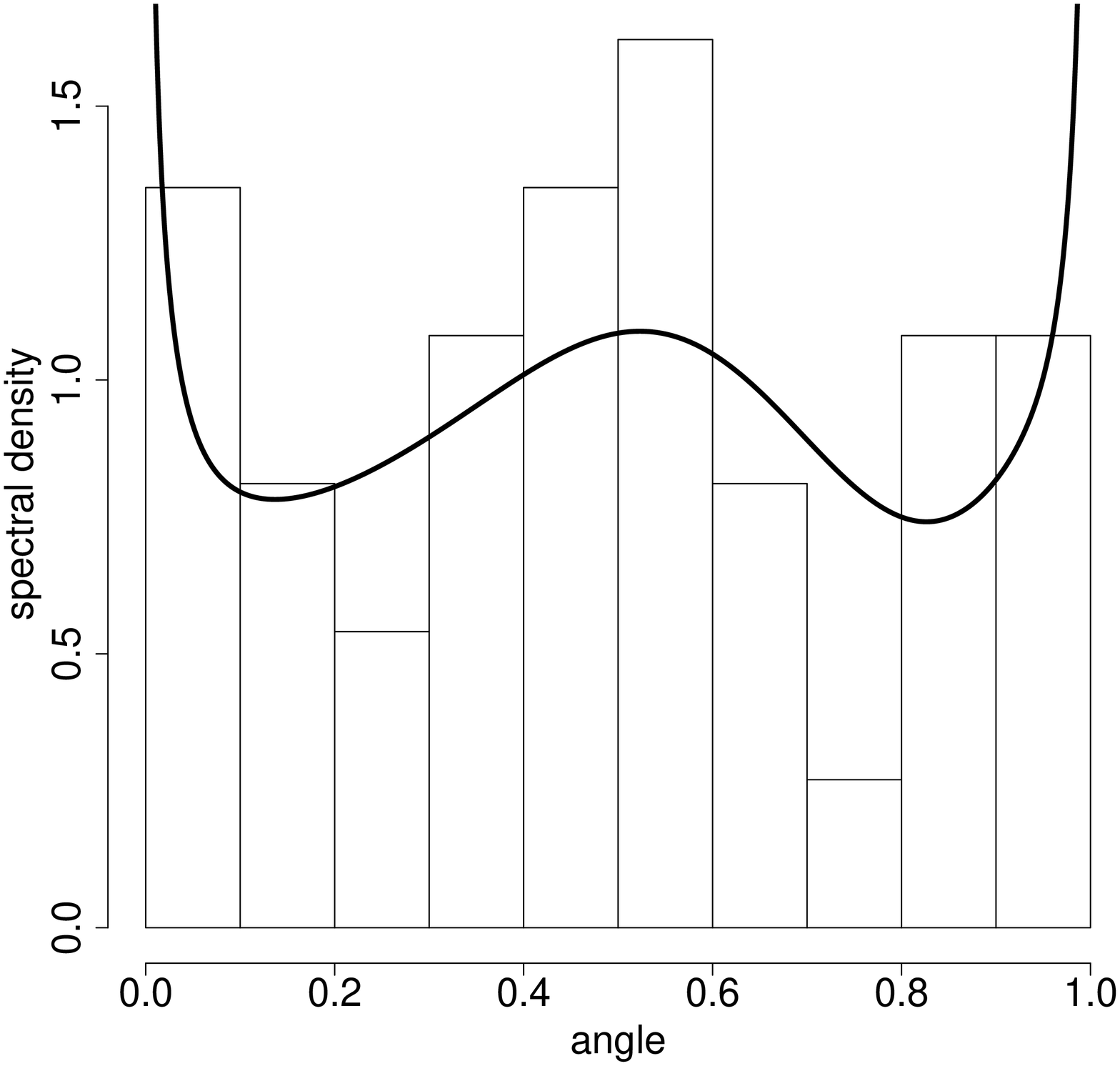}} \\
\subfigure{\includegraphics[width=0.32\textwidth]{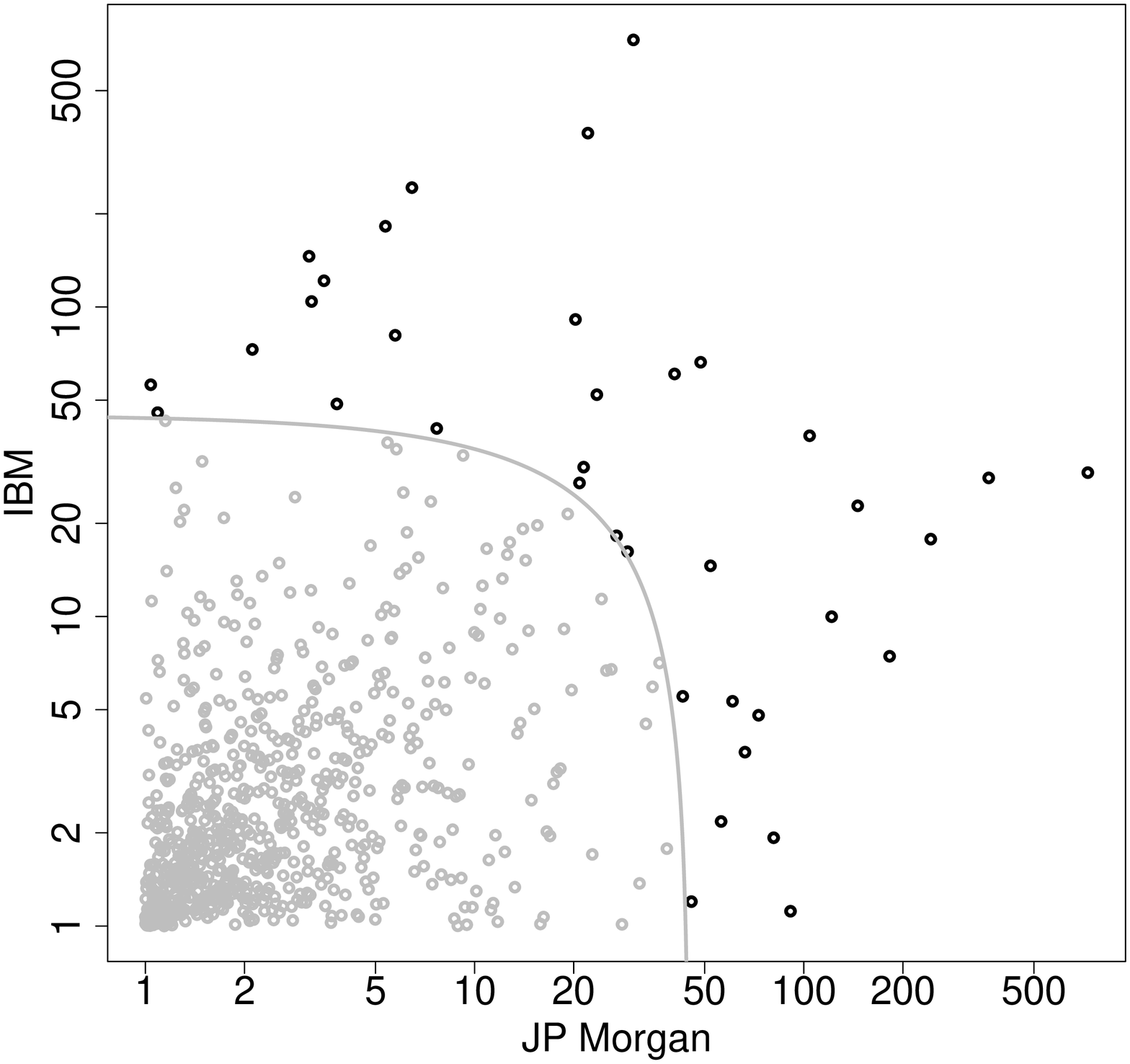}}
\subfigure{\includegraphics[width=0.32\textwidth]{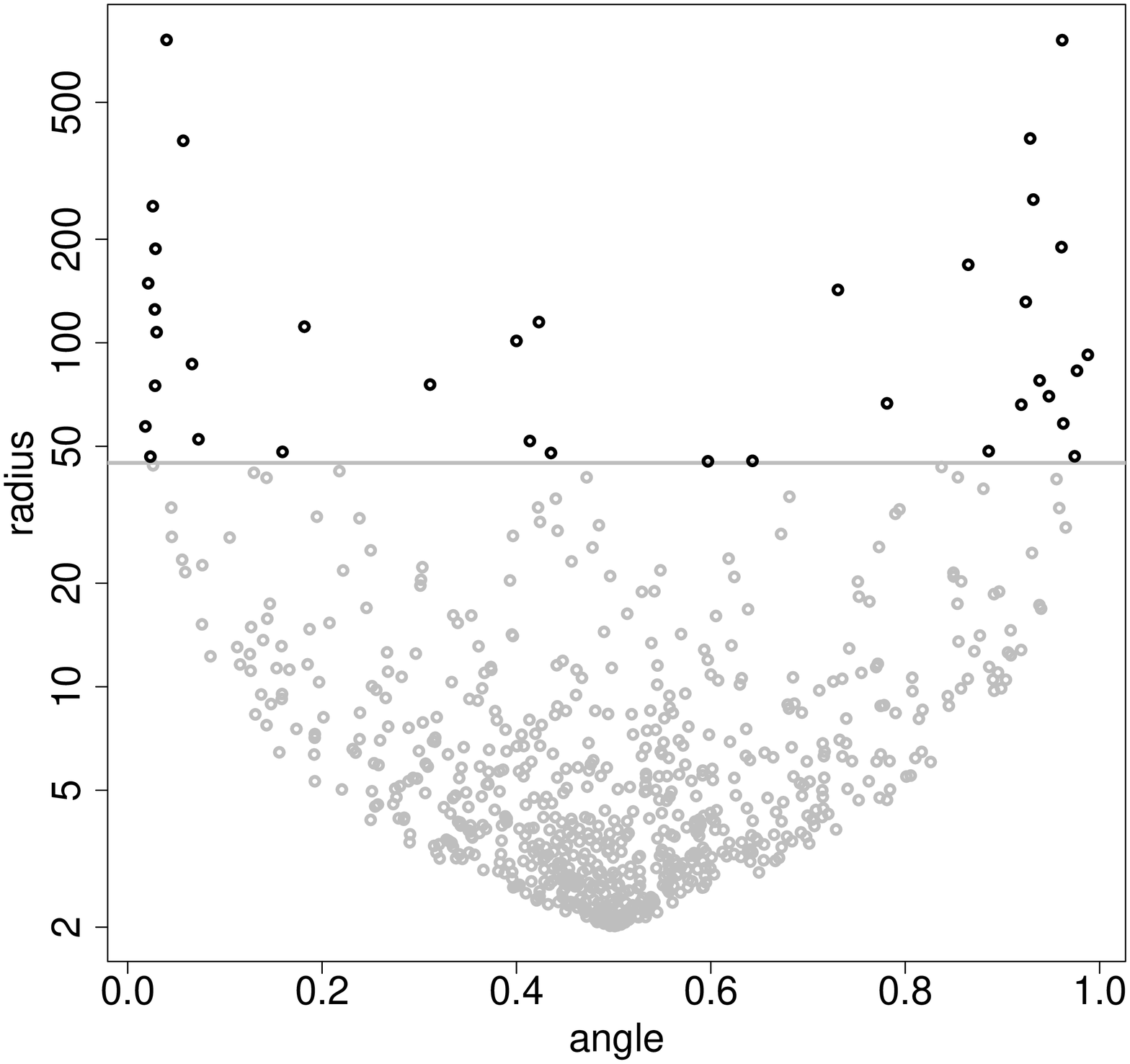}}  
\subfigure{\includegraphics[width=0.32\textwidth]{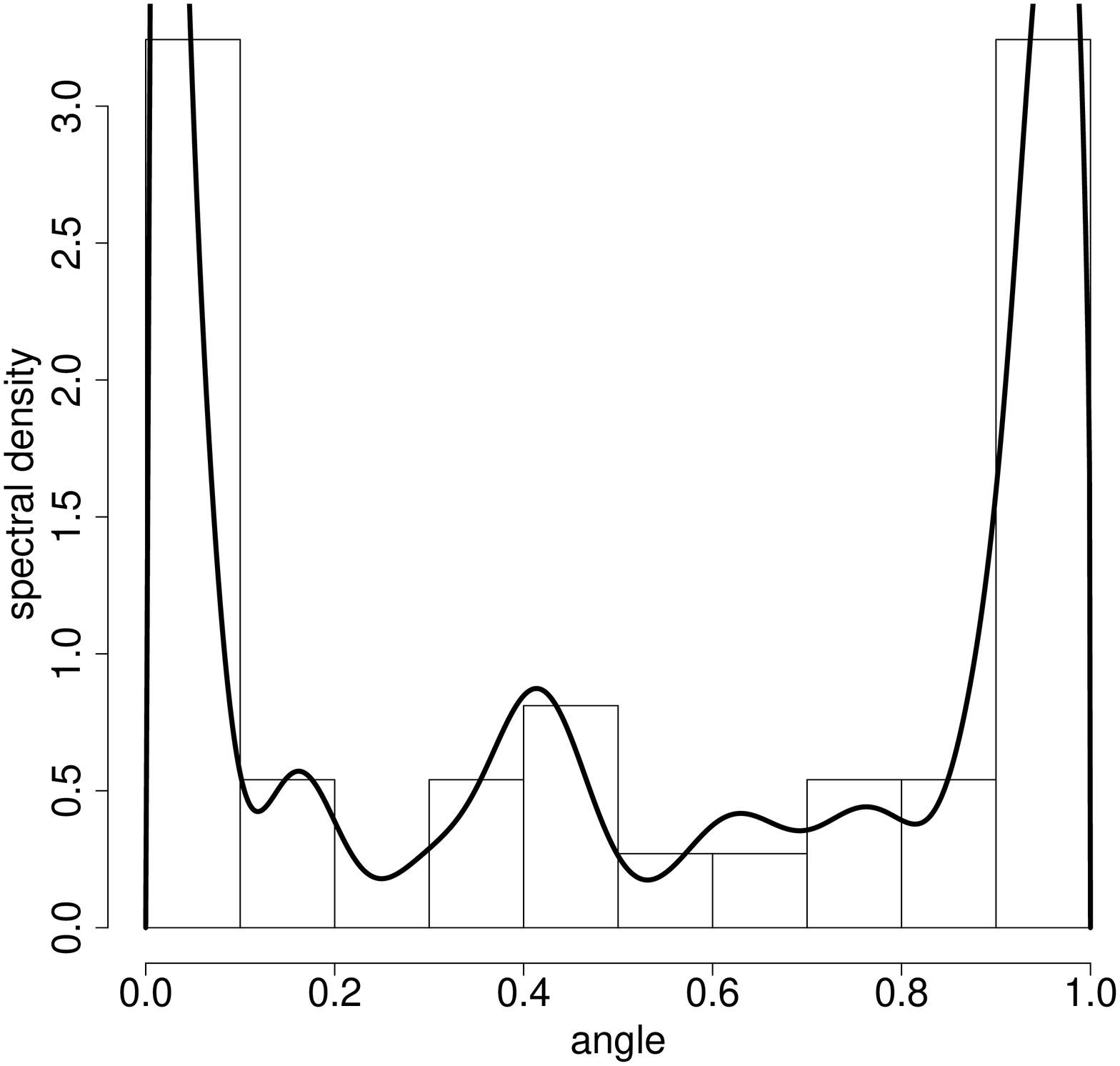}} 
\caption[Transformation into pseudo-polar coordinates for weekly stock price returns]{Log-returns of stock prices of JP Morgan versus Citibank (top) and JP Morgan versus IBM (bottom). Left: the data transformed to unit-Pareto margins (log scale); middle: further transformation into pseudo-polar coordinates; right: histogram of the angles of the extreme pairs and kernel density estimator (solid line).}
\label{fig:spec:measure2}
\end{figure}

In the general, $d$-dimensional setting we transform the random vector $\vc{X}^* = (X_{1}^*, \ldots, X_{d}^*) = (1 / \{1 - F_j (X_{j})\} : j=1,\ldots,d)$ as follows:
\begin{equation}
\label{eq:rhow}
  S = X_1^* + \cdots + X_d^*
  \quad \text{ and }
  \vc{W} = (X_{1}^*/S, \ldots, X_{d}^*/S).
\end{equation}
The spectral measure is defined as the asymptotic distribution of the vector of ratios $\vc{W}$ given that the sum $S$ is large:
\begin{equation}
\label{eq:spectral:measure1}
  \Pr \left[ \vc{W}\in \cdot \mid S > u \right]
  \dto
  H\left(\, \cdot \, \right), \qquad \text{ as } u \rightarrow \infty.
\end{equation}
The support of $H$ is included in the unit simplex $\Delta_{\dm-1} = \{ \vc{w} \in [0, \infty)^\dm : w_1 + \cdots + w_\dm = 1 \}$. In case of asymptotic independence, the spectral measure is concentrated on the $d$ vertices of $\Delta_{\dm-1}$, whereas in case of complete asymptotic dependence it reduces to a unit point mass at the center of $\Delta_{\dm-1}$. 

The spectral measure is related to the stable tail dependence function via
\begin{equation}
\label{eq:stdf_sm}
\ell(\vc{x}) = \dm\int_{\Delta_{\dm-1}} \max_{j=1,\ldots,\dm} \left\{ w_j x_j \right\} \, H(\diff \vc{w}).
\end{equation}
\index{stable tail dependence function}
The property that $\ell(0, \ldots, 0, 1, 0, \ldots, 0) = 1$ implies the moment constraints 
\begin{equation}
\label{eq:constraint1}
  \int_{\Delta_{\dm -1}} w_j \, H (\diff \vc{w}) = \frac{1}{\dm}, \qquad j=1,\ldots,\dm.
\end{equation}
Taking these constraints into account helps to improve the efficiency of nonparametric estimators of $H$ (Section~\ref{ss:estim:H}).

Any probability measure on the unit simplex satisfying \eqref{eq:constraint1} can be shown to be a valid spectral measure. 
The family of stable tail dependence functions is then characterized by~\eqref{eq:stdf_sm} via a certain integral transform of such measures.


\section{Estimation}
\label{s:estim}
\subsection{Estimating the spectral measure}
\label{ss:estim:H}

Let $(X_{i},Y_{i})$, $i=1,\ldots,n$, be independent copies of a random vector $(X,Y)$ with distribution function $F$ and continuous marginal distribution functions $F_X$ and $F_Y$. Assume that $F$ has a stable tail dependence function $\ell$ as in \eqref{eq:ellbivar}. The function $\ell$ can be represented by the spectral measure $H$ via \eqref{eq:stdf_sm}, which in the bivariate case specializes to
\begin{equation}
\label{eq:stdf_sm:biv}
  \ell(x, y) = 2 \int_0^1 \max\{ wx, (1-w)y\} \, H(\diff w), 
  \qquad (x, y) \in [0, \infty)^2,
\end{equation}
the unit simplex in $\mathbb{R}^2$ being identified with the unit interval, $[0, 1]$. The moment constraints in \eqref{eq:constraint1} simplify to
\begin{equation}
\label{eq:constraint:biv}
  \int_0^1 w \, H(\diff w) = \int_0^1 (1-w) \, H(\diff w) = \frac{1}{2}.
\end{equation}
\index{spectral measure}

The aim here is to estimate the spectral measure $H$. The starting point is the limit relation \eqref{eq:spectral:measure1}. Standardize the margins to the unit-Pareto distribution via $X^* = 1 / \{ 1 - F_X(X) \}$ and $Y^* = 1 / \{ 1 - F_Y(Y) \}$ and write
\begin{align*}
  S &= X^* + Y^*, & 
  W &= X^* / (X^* + Y^*).
\end{align*}
Equation~\eqref{eq:spectral:measure1} then becomes
\begin{equation}
\label{eq:spectral:measure:biv}
  \Pr[ W \in \cdot \mid S > u ] \dto H(\,\cdot\,), 
  \qquad \text{ as } u \to \infty.
\end{equation}
The spectral measure $H$ can be interpreted as the limit distribution of the angular component, $W$, given that the radial component, $S$, is large.

Mimicking the transformation from $(X, Y)$ via $(X^*, Y^*)$ to $(S, W)$ in a given data-set, it is convenient to standardize the data to the unit-Pareto distribution via \eqref{eq:transPareto} and then switch to pseudo-polar coordinates using \eqref{eq:transPolar}. Different choices for the angular components are possible; for example, \cite{einmahl2009} use $\arctan(X^* / Y^*)$. For the radial component one can take in general the $L_p$ norm, $p \in [1, +\infty]$; \cite{einmahl2001} take the $L_\infty$ norm and \cite{carvalho2013} take the $L_1$ norm. Note however that different normalizations lead to different spectral measures. 

For estimation purposes, transform pairs $(X_i, Y_i)$ into $(\widehat{S}_i, \widehat{W}_i)$ as in \eqref{eq:transPareto} and \eqref{eq:transPolar}. Set the threshold $u$ in \eqref{eq:spectral:measure:biv} to $n/k$, where $k=k_n\in(0,n]$ is an intermediate sequence such that $k\to\infty$ and $k/n\to 0$ as $n\to \infty$. Let $I_n=\{i=1,\ldots,n:\;\widehat{S}_i\geq n/k\}$ denote the set of indices that correspond to the observations with pseudo-radial component $\widehat{S}_i$ above a high threshold, and let $N_n = \abs{ I_n }$ denote its cardinality. Here we specify $u$ as the $95\%$ quantile of $\widehat{S}_i$.
\index{spectral measure}

In the literature, three nonparametric estimators of the spectral measure have been proposed. All three are of the form 
\begin{equation}
\label{eq:spectral:measure:est1}
\widehat{H}_{l}\left(x\right)
=\sum_{i\in I_n}\widehat{p}_{l,i} \, \1 \{\widehat{W}_i\leq x\}, 
\qquad x \in [0, 1],\quad l=1,2,3.
\end{equation} 
The estimators distinguish themselves in the way the weights $\widehat{p}_{l,i}$ are defined.

For the \emph{empirical spectral measure} \citep{einmahl2001}, the weights are set to be constant, i.e., $\widehat{p}_{1,i}=1/N_n$. The estimator $\widehat{H}_1$  \eqref{eq:spectral:measure:est1} becomes an empirical version of \eqref{eq:spectral:measure:biv}. However, this estimator does not necessarily satisfy the moment constraints in \eqref{eq:constraint:biv}. This is a motivation for the two other estimators where the moment constraint is enforced by requiring that $\sum_{i\in I_n} \widehat{W}_i \widehat{p}_{l,i} = 1/2$.
\index{spectral measure!empirical spectral measure} 

The \emph{maximum empirical likelihood estimator} \citep{einmahl2009}, $\widehat{H}_{2}\left(x\right)$, has probability masses $\widehat{p}_{2,i}$ solving the optimization problem
\begin{equation} \label{optim_mele}
  \begin{array}{rl}
    \underset{\bm{p} \in \mathbb{R}^k_+}{\max}        &\sum_{i\in I_n} \log p_i \\
    \mbox{s.t.} & \sum_{i\in I_n} p_i = 1 \\
                & \sum_{i\in I_n} \widehat{W}_i p_i = 1/2.
    \end{array}
\end{equation}
\index{spectral measure!maximum empirical likelihood estimator}%
By construction, the weights $\widehat{p}_{2,i}$ satisfy the moment constraints. The optimization problem in \eqref{optim_mele} can be solved by the method of Lagrange multipliers. In the nontrivial case where $1/2$ is in the convex hull of $\{W_i : i \in I_n\}$, the solution is given by
\[
   \widehat{p}_{2,i} = \frac{1}{N_n} \frac{1}{1+ \lambda (\widehat{W}_i-1/2)}, \qquad i\in I_n,
\]
where $\lambda \in \mathbb{R}$ is the Lagrange multiplier associated to the second equality constraint in \eqref{optim_mele}, defined implicitly as the solution to the equation
\[
  \frac{1}{N_n} \sum_{i\in I_n} \frac{\widehat{W}_i-1/2}{{1+ \lambda (\widehat{W}_i-1/2)}}=0.
\]
Note that in \eqref{optim_mele}, it is implicitly assumed that  $\widehat{p}_{2,i}>0$.

Another estimator that satisfies the moment constraints is the \emph{maximum Euclidean likelihood estimator}
\index{spectral measure!maximum Euclidean likelihood estimator} 
\citep{carvalho2013}. The probability masses $\widehat{p}_{3,i}$ solve the following optimization problem:
\begin{equation}
  \begin{array}{rl}
    \underset{\bm{p} \in \mathbb{R}^{I_n}}{\max}&-\frac{1}{2} \sum_{i\in I_n} (N_n p_i - 1)^2 \\
    \text{s.t.} & \sum_{i\in I_n} p_i = 1 \\
                & \sum_{i\in I_n} \widehat{W}_i p_i = 1/2.
  \end{array}
\end{equation}
This quadratic optimization problem with linear constraints can be solved explicitly with the method of Lagrange multipliers, yielding
\begin{equation}
  \widehat{p}_{3,i} = \frac{1}{N_n} \bigl\{ 1- (\overline{W} - 1/2)S_W^{-2} (\widehat{W}_i - \overline{W}) \bigr\}, \qquad i\in I_n,
\end{equation}
where $\overline{W}$ and $S_{W}^2$ denote the sample mean and sample variance of $\widehat{W}_i$, $i\in I_n$, respectively, that is,
\begin{align*}
  \overline{W} &= \frac{1}{N_n} \sum_{i\in I_n} \widehat{W}_i, & 
  S_{W}^2 &= \frac{1}{N_n} \sum_{i\in I_n} (\widehat{W}_i - \overline{W})^2.
\end{align*}
The weights $\widehat{p}_{3,i}$ could be negative, but this does usually not occur as the weights are all nonnegative with probability tending to one.

It is shown in \cite{einmahl2009} and \cite{carvalho2013} that $\widehat{H}_{2}\left(x\right)$ and $\widehat{H}_{3}\left(x\right)$ are more efficient than $\widehat{H}_{1}\left(x\right)$. Moreover, asymptotically there is no difference between the maximum empirical likelihood or maximum Euclidean likelihood estimators. The maximum Euclidean likelihood estimator is especially convenient as the weights $\widehat{p}_{3,i}$ are given explicitly. 

For the stock market data from JP Morgan, Citibank and IBM, Figure~\ref{fig:spec:measure3} shows the empirical spectral measure $\widehat{H}_{1}\left(x\right)$ and the maximum Euclidean estimator $\widehat{H}_{3}\left(x\right)$ for the two pairs involving JP Morgan. In each case, the threshold $u$ is set to be the $95\%$ quantile of the sample of radii $\widehat{S}_i$. Enforcing the moment constraints makes a small but noticeable difference. Tail dependence is strongest for the pair JP Morgan versus Citibank, with spectral mass distributed approximately uniformly over $[0, 1]$. For the pair JP Morgan versus IBM, tail dependence is weaker, the spectral measure being concentrated mainly near the two endpoints, $0$ and $1$.

The nonparametric estimators of the spectral probability measure can be smoothed in such a way that the smoothed versions still obey the moment constraints. This can be done using kernel smoothing techniques, although some care is needed since the spectral measure is defined on a compact interval. In \cite{carvalho2013} the estimator is constructed by combining Beta distributions with the weights, $\widehat{p}_{3,i}$, of the maximum Euclidean likelihood estimate. To ensure that the estimated measure obeys the marginal moment constraints, it is imposed that the mean of each smoother equals the observed pseudo-angle. The Euclidean spectral density estimator is defined as
\begin{equation} \label{h.smooth}
  \widetilde{h}_3 (x) 
  = \sum_{i\in I_n} \widehat{p}_{3,i} \, 
  \beta \bigl(x;  \widehat{W}_i \nu, (1- \widehat{W}_i)\nu \bigr), \qquad x \in (0,1),
\end{equation}
where $\nu>0$ is the smoothing parameter, to be chosen via cross-validation, and
\[
  \beta\left(x; p, q\right)
  =\left\{  \int_{0}^1 u^{p-1} (1-u)^{q-1} \, \diff u\right\}^{-1} 
  x^{p-1}(1-x)^{q-1} \, \mathbbm{1}_{[0, 1]}(x),
\]
is the Beta density with parameters $p, q > 0$. 

For the financial data, the realized spectral density estimator $\widetilde{h}_3 (x)$ is plotted on the right-hand panel of Figure~\ref{fig:spec:measure2}. The picture is in agreement with the one in Figure~\ref{fig:spec:measure3}. The distribution of the spectral mass points to asymptotic dependence for JP Morgan versus Citibank and to weaker asymptotic dependence for JP Morgan versus IBM.

Integrating out the estimated spectral density yields a smoothed version of the empirical spectral measure or its variants: for the maximum Euclidean likelihood estimator, we get
\begin{equation} 
  \widetilde{H}_{3}(x) 
  = \int_0^x \widetilde{h}_{3}(v) \, \diff v 
  = \sum_{i\in I_n} \widehat{p}_{3,i} \,
  \mathcal{B} \bigl(x;  \widehat{W}_i \nu, (1- \widehat{W}_i) \nu\bigr), \qquad x \in [0,1],
\end{equation}
with $\mathcal{B}(x;p,q) = \int_{0}^{x} \beta(y; p, q) \, \diff y$ the regularized incomplete beta function. As
\begin{equation}
  \int_0^1 x \, \widetilde{h}_3(x) \, \diff x 
  = \sum_{i\in I_n} \widehat{p}_{3,i}
  \, \frac{\nu  \widehat{W}_i}{\nu  \widehat{W}_i + \nu(1- \widehat{W}_i)}
  = \sum_{i\in I_n} \widehat{p}_{3,i}  \widehat{W}_i = 1/2,
\end{equation}
the moment constraint \eqref{eq:constraint:biv} is satisfied.

\begin{figure}[ht]
\centering
\subfigure{\includegraphics[width=0.48\textwidth]{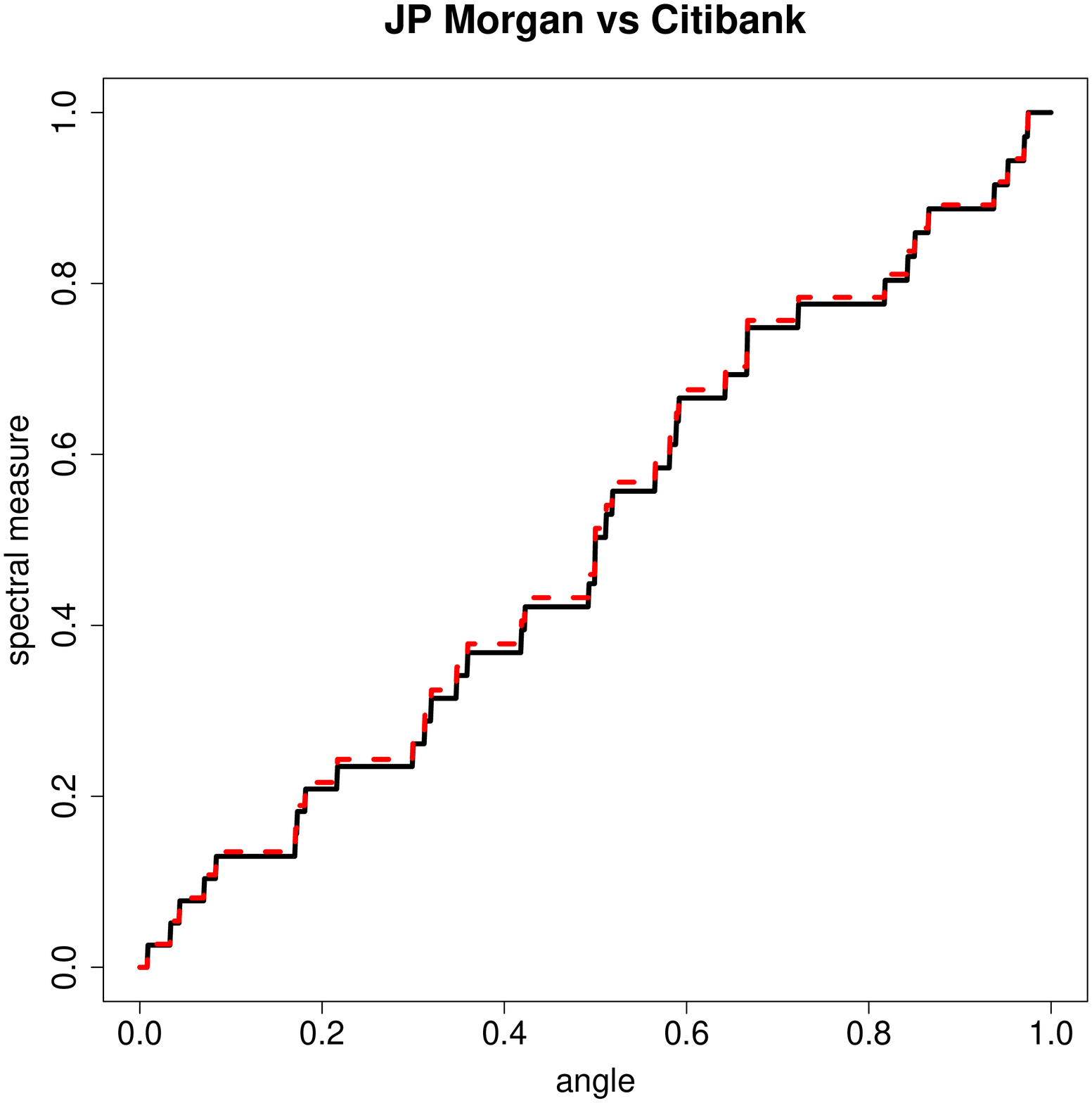}}  
\subfigure{\includegraphics[width=0.48\textwidth]{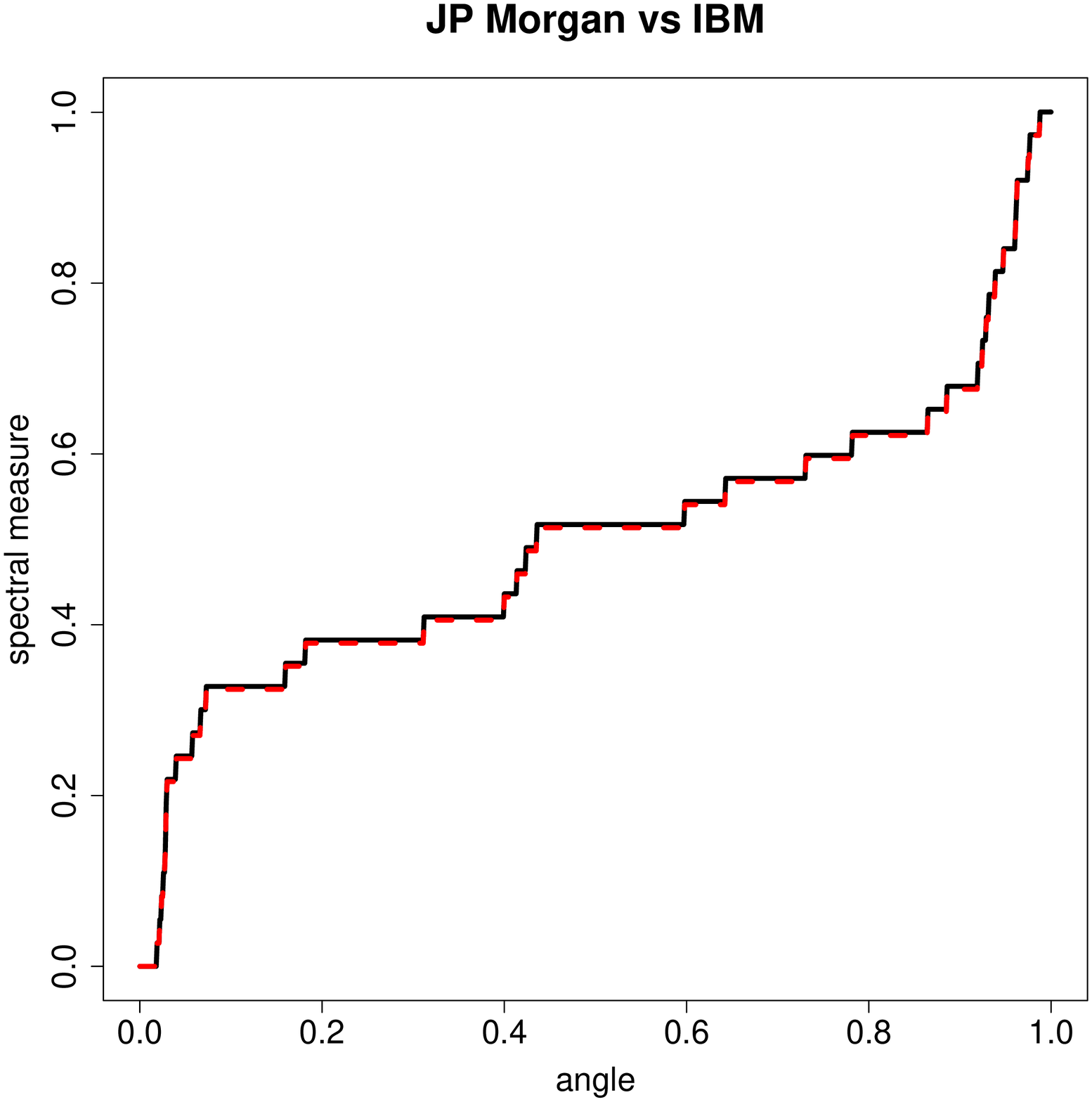}}  
\caption[Empirical spectral measures for the weekly stock price returns]{Estimated empirical spectral measures for JP Morgan versus Citibank (left) and JP Morgan versus IBM (right). The solid line and the dashed line correspond to $\widehat{H}_{3}(x)$ and $\widehat{H}_{1}(x)$ respectively.}
\label{fig:spec:measure3}
\end{figure}

\subsection{Estimating the stable tail dependence function}
\label{ss:estim:stdf}

Consider again the general, multivariate case. Given a random sample $\vc{X}_i = (X_{i1}, \ldots, X_{id})$, $i = 1, \ldots, n$, the aim is to estimate the stable tail dependence function $\ell$ in \eqref{eq:ell} of the common distribution $F$ with continuous margins $F_1, \ldots, F_d$. A straightforward nonparametric estimator can be defined as follows. Let $k=k_n \in \{1, \ldots , n\}$ be such that $k \to \infty$ and $k/n \to 0$ as $n \to \infty$. By replacing $\mathbb{P}$ by the empirical distribution function, $t$ by $k/n$, and $F_1,\ldots,F_d$ by $\widehat{F}_1,\ldots,\widehat{F}_d$ as defined in \eqref{eq:empiric}, we obtain the \emph{empirical tail dependence function} \citep{huang1992, drees1998}
\begin{align*}
  \widehat{\ell} (\vc{x}) 
  &\coloneqq \frac{n}{k} \frac{1}{n} \sum_{i=1}^n 
  \1 \left\{ 
    \widehat{F}_1 (X_{i1}) > 1 - \frac{k x_1}{n} \text{ or } 
    \ldots \text{ or } 
    \widehat{F}_d (X_{id}) > 1 - \frac{k x_d}{n} 
    \right\} \\
  &=  \frac{1}{k} \sum_{i=1}^n 
  \1 \left\{ 
    R_{i1,n} > n + 1 - k x_1 \text{ or } \ldots \text{ or } R_{id,n} > n + 1 - k x_d 
  \right\}.
\end{align*}
\index{stable tail dependence function}%
\index{stable tail dependence function!empirical stable tail dependence function}%
Under minimal assumptions, the estimator is consistent and asymptotically normal with a convergence rate of $\sqrt{k}$ \citep{einmal:krajina:segers:2012, buecher:segers:volgushev:2014}. Alternatively, the marginal distributions might be estimated by $\widehat{F}_{j,2} (X_{ij}) = R_{ij,n}/ n$ or $\widehat{F}_{j,3} (X_{ij}) = (R_{ij,n} - 1/2)/n$, resulting in estimators that are asymptotically equivalent to $\widehat{\ell}$.

Another estimator of $\ell$ can be defined by estimating the spectral measure, $H$, and applying the transformation in \eqref{eq:stdf_sm}, an idea going back to \citet{caperaa2000}. In the bivariate case, we can replace $H$ in \eqref{eq:stdf_sm:biv} by one of the three nonparametric estimators $\widehat{H}_l$ studied in Section~\ref{ss:estim:H}.
Recall the notations in \eqref{eq:transPareto} and \eqref{eq:transPolar}. If we modify the definition of the index set $I_n$ to $I_n = \{i=1,\ldots,n : \widehat{S}_i > \widehat{S}_{(k+1)} \}$, where $\widehat{S}_{(k+1)}$ denotes the $(k+1)$-th largest observation of the $\widehat{S}_i$'s, then there are exactly $k$ elements in the set $I_n$. Starting from the empirical spectral measure $\widehat{H}_1$, we obtain the estimator
\begin{equation*}
\widehat{\ell}_{CF} (t) 
= \frac{1}{k} \sum_{i \in I_n} \max \left\{ \widehat{W}_i x, (1-\widehat{W}_i) y \right\}.
\end{equation*}
For the maximum empirical or Euclidean likelihood estimators $\widehat{H}_{2}$ and $\widehat{H}_{3}$, 
one needs to replace the factor $1/k$ by  the weights $\widehat{p}_{2,i}$ and $\widehat{p}_{3,i}$, respectively.

A way to visualize the function $\ell$ or an estimator thereof is via the level sets $\mathcal{D}_c \coloneqq \{(x,y) : \ell(x,y) = c \}$ for a range of value of $c > 0$. Note that the level sets are equal to the lines $x + y = c$ in case of asymptotic independence and to the elbow curves $\max(x, y) = c$ in case of complete asymptotic dependence.
Likewise, a plot of the level sets of an estimator of $\ell$ can be used as a graphical diagnostic of asymptotic (in)dependence; see \citet{dehaan1998} or \citet[Section~7.2]{dehaan2006}.

We plot the lines $\mathcal{D}_c$ for $c \in \{0.2,0.4,0.6,0.8,1\}$ and $k = 50$ of $\widehat{\ell}$ and $\widehat{\ell}_{CF}$ for the weekly log-returns of JP Morgan versus Citibank and JP Morgan versus IBM in Figure~\ref{fig:elllevel}. The level sets for JP Morgan versus IBM resemble the straight lines $x + y = c$ much more closely than the level sets for JP Morgan versus Citibank do. The estimator based on the spectral measure, $\widehat{\ell}_{CF}$, acts as a smooth version of the empirical tail dependence function $\widehat{\ell}$.

\begin{figure}[ht]
\centering
\subfigure{\includegraphics[width=0.48\textwidth]{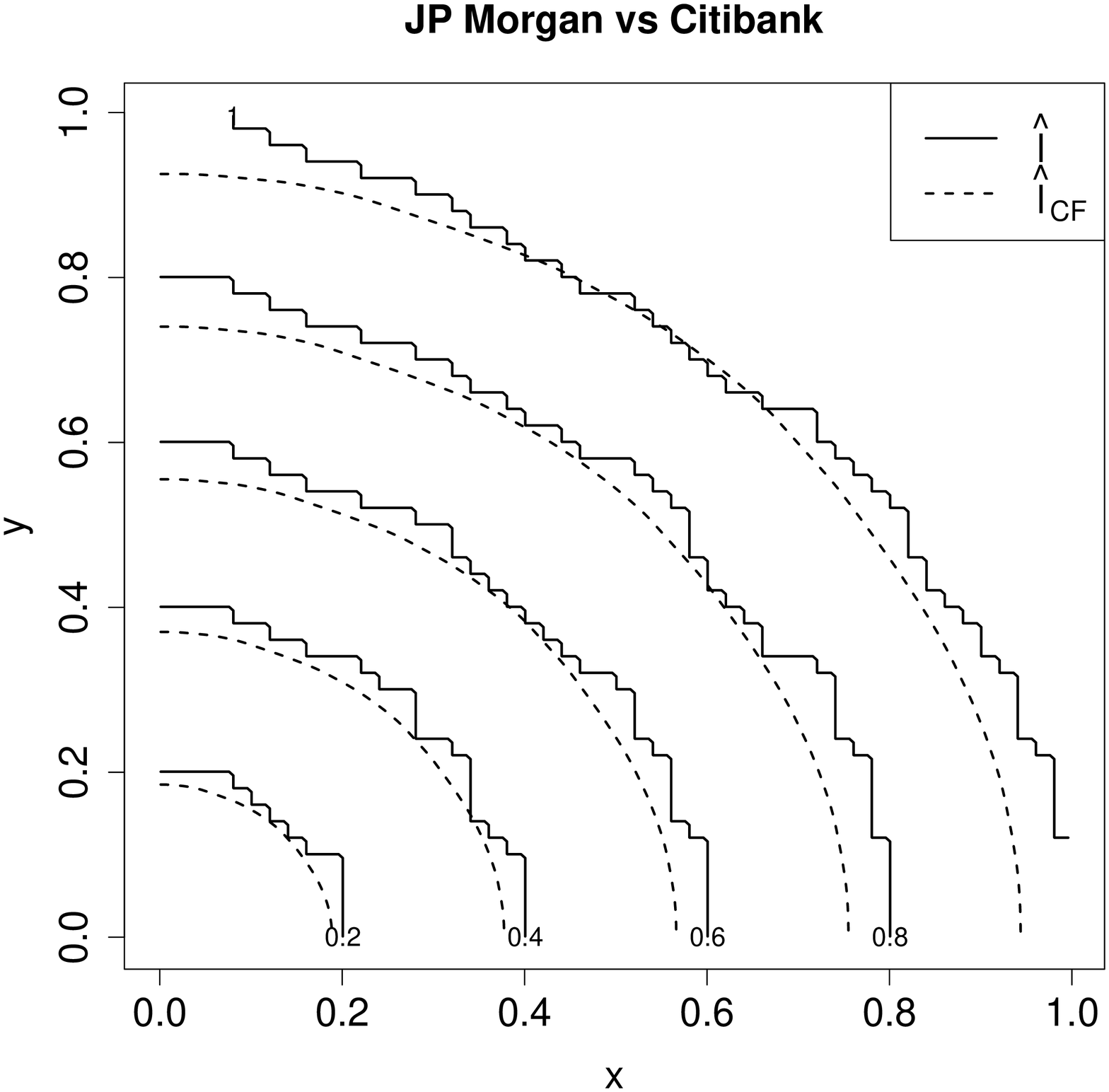}}
\subfigure{\includegraphics[width=0.48\textwidth]{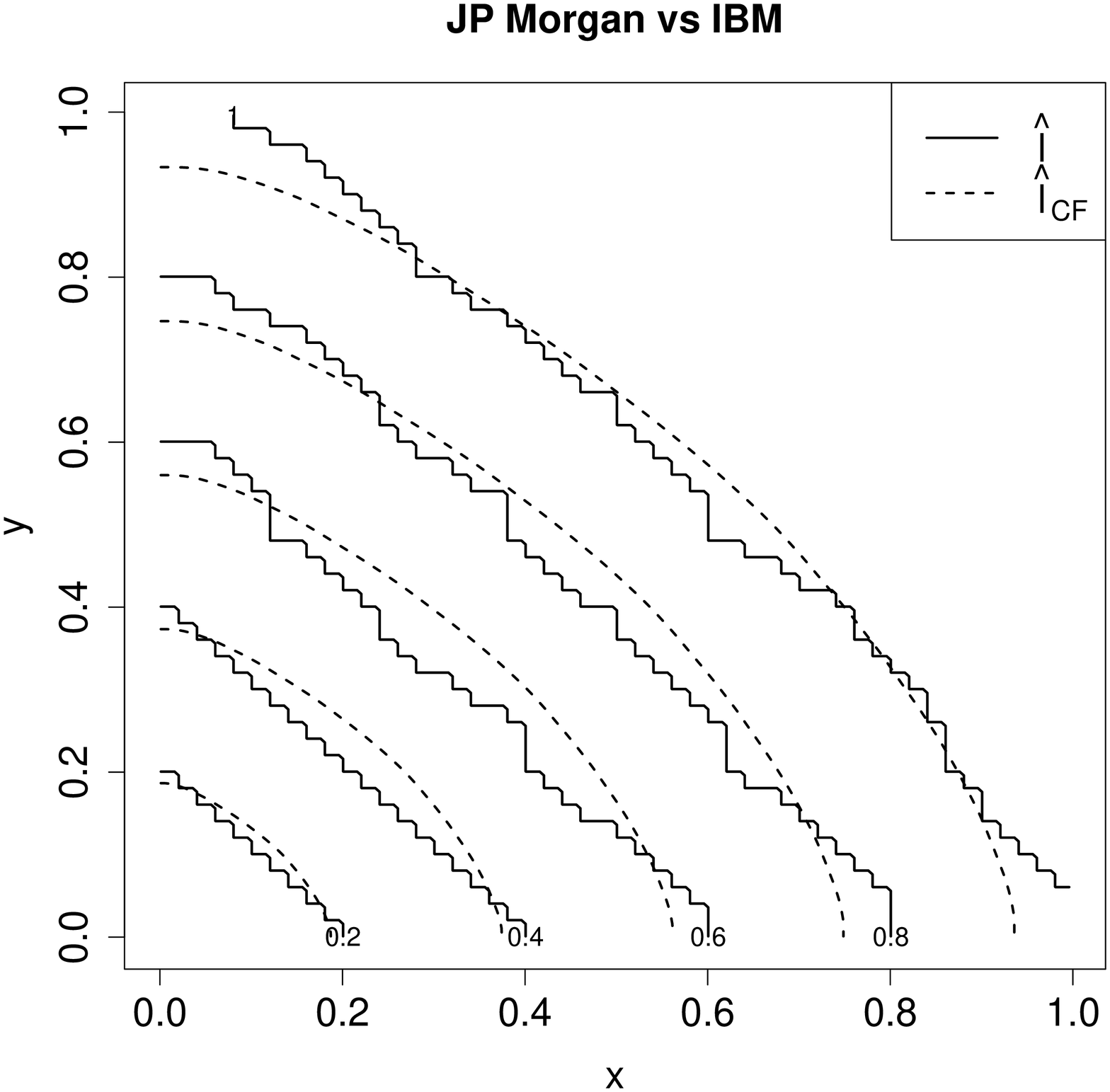}}     
\caption[Level sets of the stable tail dependence function $\ell$]{Level sets $\mathcal{D}_c$ of $\widehat{\ell} (x,y)$ and $\widehat{\ell}_{CF}$ for $c \in \{ 0.2, 0.4, 0.6, 0.8, 1 \}$.}
\label{fig:elllevel}
\end{figure}

Nonparametric estimators for $\ell$ can also serve as stepping stone for semiparametric inference. Assume that $\ell = \ell_{\theta_0} \in \{ \ell_\theta : \theta \in \Theta \}$, a finite-dimensional parametric family. Then one could estimate $\theta_0$ by minimizing a distance or discrepancy measure between $\hat{\ell}$ and members of the parametric family \citep{einmal:krajina:segers:2012,einmahl2014}.

\section{Asymptotic independence}
\label{s:AI}
In this section we will focus on the bivariate case. 
Let $(X_i,Y_i)$, $i=1,\ldots,n$, be independent copies of a random vector $(X,Y)$, with distribution function $F$ having continuous margins $F_X$ and $F_Y$. Recall from~\eqref{eq:ell} that the stable tail dependence function (provided it exists) is given by
\begin{equation}
\label{eq:ellbivar}
  \ell(x,y) = 
  \lim_{t \downarrow 0} t^{-1} \, 
  \Pr[ 1 - F_X (X) \leq xt \text{ or } 1 - F_Y (Y) \leq yt],
  \qquad (x, y) \in [0, \infty)^2.
\end{equation}
After marginal standardization, the pair of component-wise sample maxima converges weakly to the bivariate max-stable distribution $G_0$ given by $G_0(z_1, z_2) = \exp\{ - \ell(1/z_1, 1/z_2) \}$ for $(z_1, z_2) \in (0, \infty)^2$; see \eqref{eq:tail}. \emph{Asymptotic independence} of the sample maxima, $G_0(z_1, z_2) = \exp(-1/z_1) \exp(-1/z_2)$ for all $(z_1, z_2)$, is equivalent to $\ell(x, y) = x + y$ for all $(x, y)$. The opposite case, $\ell(x, y) < x+y$ for some $(x, y)$, is referred to as \emph{asymptotic dependence}.
\index{asymptotic independence}

\citet{sibuya1960} already observed that bivariate normally distributed vectors are asymptotically independent as soon as the correlation is less than unity. We find that in case of asymptotic independence, we cannot rely on the function $\ell$ to quantify the amount of dependence left above high but finite thresholds. 

Tail dependence coefficients serve to quantify the amount of tail dependence and to distinguish between asymptotic dependence versus independence (Subsection~\ref{ss:ai:coef}). To decide which of the two situations applies for a data set, a number of testing procedures are available (Subsection~\ref{ss:ai:test}).

\subsection{Tail dependence coefficients}
\label{ss:ai:coef}

A first measure for the strength of asymptotic dependence is the \emph{tail dependence coefficient} \index{tail dependence coefficient}
\begin{equation}
\label{eq:chi}
  \chi 
  \coloneqq
  \lim_{u \uparrow 1} \Pr[F_X (X) > u \mid F_Y (Y) > u]
  = R(1,1) = 2 - \ell(1, 1)
\end{equation}
\citep{coles1999}. By the properties of $\ell$ (Section~\ref{ss:prob:stdf}), we have $\ell(x, y) = x + y$ for all $(x, y)$ if and only if $\ell(1, 1) = 2$. That is, asymptotic independence is equivalent to $\chi = 0$, whereas asymptotic dependence is equivalent to $\chi \in (0, 1]$. 

To estimate $\chi$, we will first write it as the limit of a function $\chi (u)$, so that $\lim_{u \uparrow 1} \chi (u) = \chi$; the function $\chi (u)$ is defined as
\begin{equation*}
  \chi(u) 
  \coloneqq  2 - \frac{1 -  \PP[F_X (X) < u, \, F_Y (Y) < u]}{1-u}, 
  \qquad 0 \leq u < 1.
\end{equation*}
To estimate $\chi(u)$ from a sample $(X_1, Y_1), \ldots, (X_n, Y_n)$, simply replace $F_X$ and $F_Y$ by their empirical counterparts in \eqref{eq:empiric}: 
\[
  \widehat{\chi}(u) 
  \coloneqq 2 - \frac{1}{1-u} 
  \biggl( 
      1 - \frac{1}{n} \sum_{i=1}^n 
      \mathbbm{1} \left( \widehat{F}_X(X_i) < u, \, \widehat{F}_Y(Y_i) < u \right) 
  \biggr).
\]

For the stock price log-returns of JP Morgan, Citibank, and IBM, the estimated tail dependence coefficients $\widehat{\chi}(u)$ of the three possible pairs are shown in Figure~\ref{fig:chiplot} as a function of $u \ge 0.8$.
The plot confirms the earlier finding that tail dependence is stronger between the two banking stocks, JP Morgan and Citibank, than between either of the two banks and IBM.

If the variables are asymptotically independent, $\chi = 0$, we need a second measure quantifying the extremal dependence. The \emph{coefficient of tail dependence}, $\eta$, can be motivated as follows \citep{ledford1996}. Since $\Pr[1/\{1 - F_X (X)\} > t] = 1/t$ for $t \ge 1$, we have
\index{coefficient of tail dependence}
\begin{equation*}
\Pr \left[ \frac{1}{1 - F_X (X)} > t, \, \frac{1}{1 - F_Y (Y)} > t \right] \propto
\begin{dcases}
t^{-1} & \text{ for asymptotic dependence,} \\
t^{-2} & \text{ for exact independence.}
\end{dcases} 
\end{equation*}
A model that links these two situations is given by the assumption that the joint survival function above is regularly varying at $\infty$ with index $-1/\eta$, i.e.,
\begin{equation}
\label{eq:eta2}
\PP \left[ \frac{1}{1 - F_X (X)} > t, \frac{1}{1 - F_Y (Y)} > t \right] =  t^{-1/\eta} \, \mathcal{L}(t),
\end{equation}
where $\mathcal{L}$ is a slowly varying function, that is, $\mathcal{L} (tr)/ \mathcal{L}(t) \to 1$ as $t \to \infty$ for fixed $r > 0$; see for example \citet{resnick1987}.

For large $t$, if we treat $\mathcal{L}$ as constant, there are three cases of non-negative extremal association to distinguish:
\begin{enumerate}
\item $\eta = 1$ and $\mathcal{L} (t) \to c > 0$: asymptotic dependence;
\item $ 1/2 < \eta < 1$: positive association within asymptotic independence;
\item $\eta = 1/2$: (near) perfect independence.
\end{enumerate}
Values of $\eta < 1/2$ occur when there is negative association between the two random variables at extreme levels.

In order to estimate $\eta$, we first define the structure variable
\begin{equation}
\label{eq:Tmin}
  T \coloneqq \min \left( \frac{1}{1 - F_X (X)}, \frac{1}{1 - F_Y (Y)} \right).
\end{equation}
For a high threshold $u$ and for $t \ge 0$, we have, by slow variation of $\mathcal{L}$,
\begin{equation*}
\PP [ T > u + t \mid T > u] 
= \frac{\mathcal{L} (u + t)}{\mathcal{L} (u)}  (1 + t/u)^{-1/ \eta} 
\approx (1 + t/u)^{-1/ \eta}. 
\end{equation*}
On the right-hand side we recognize a generalized Pareto distribution with shape parameter $\eta$. Given the sample $(X_1, Y_1), \ldots, (X_n, Y_n)$, we can estimate $\eta$ with techniques from univariate extreme value theory applied to the variables
\[
  \widehat{T}_i \coloneqq 
  \min \left( 
    \frac{1}{1 - \widehat{F}_X (X_i)}, \,
    \frac{1}{1 - \widehat{F}_Y (Y_i)} 
  \right), \qquad i = 1, \ldots, n.
\]
One possibility consists of fitting the generalized Pareto distribution to the sample of excesses $\widehat{T}_i - u$ for those $i$ for which $\widehat{T}_i > u$ \citep[page 180]{ledford1996}. Alternatively, one can use the Hill estimator \citep{hill1975}
\index{Hill estimator}
\begin{equation*}
\widehat{\eta}_H 
\coloneqq \frac{1}{k} \sum_{i=1}^k \log \left( \frac{\widehat{T}_{(i)}}{\widehat{T}_{(k+1)}} \right),
\end{equation*}
where $\widehat{T}_{(1)} \geq \cdots \geq \widehat{T}_{(n)}$ denote the order statistics of $\widehat{T}_1,\ldots,\widehat{T}_n$. Figure~\ref{fig:chiplot} shows the estimators $\widehat{\eta}_H$ of the coefficient of tail dependence between JP Morgan and Citibank, JP Morgan and IBM, and Citibank and IBM for a decreasing number of order statistics. The estimators suggest asymptotic dependence for the log-returns of JP Morgan versus Citibank.

\begin{figure}[ht]
\centering
\subfigure{\includegraphics[width=0.48\textwidth]{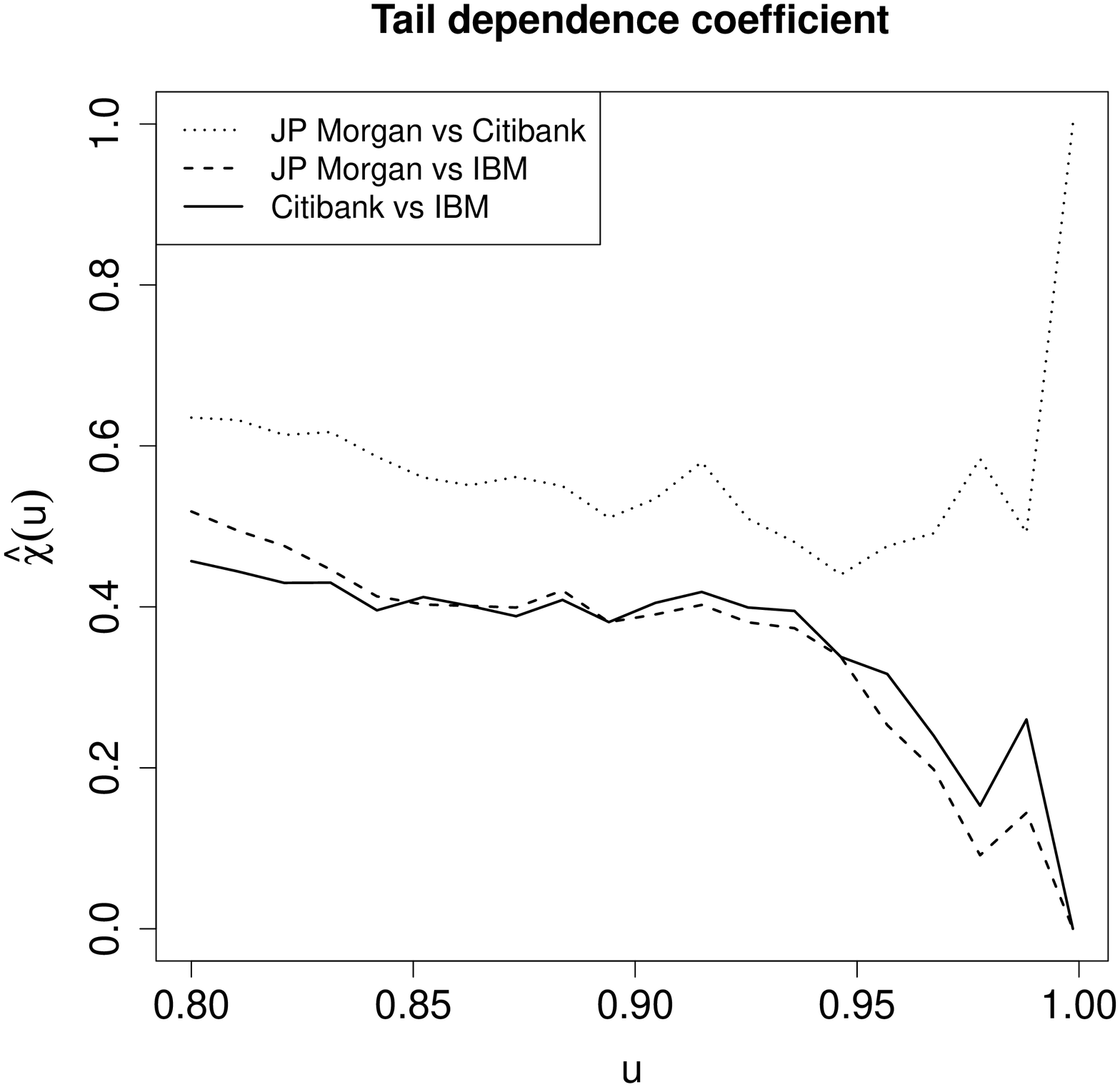}}
\subfigure{\includegraphics[width=0.48\textwidth]{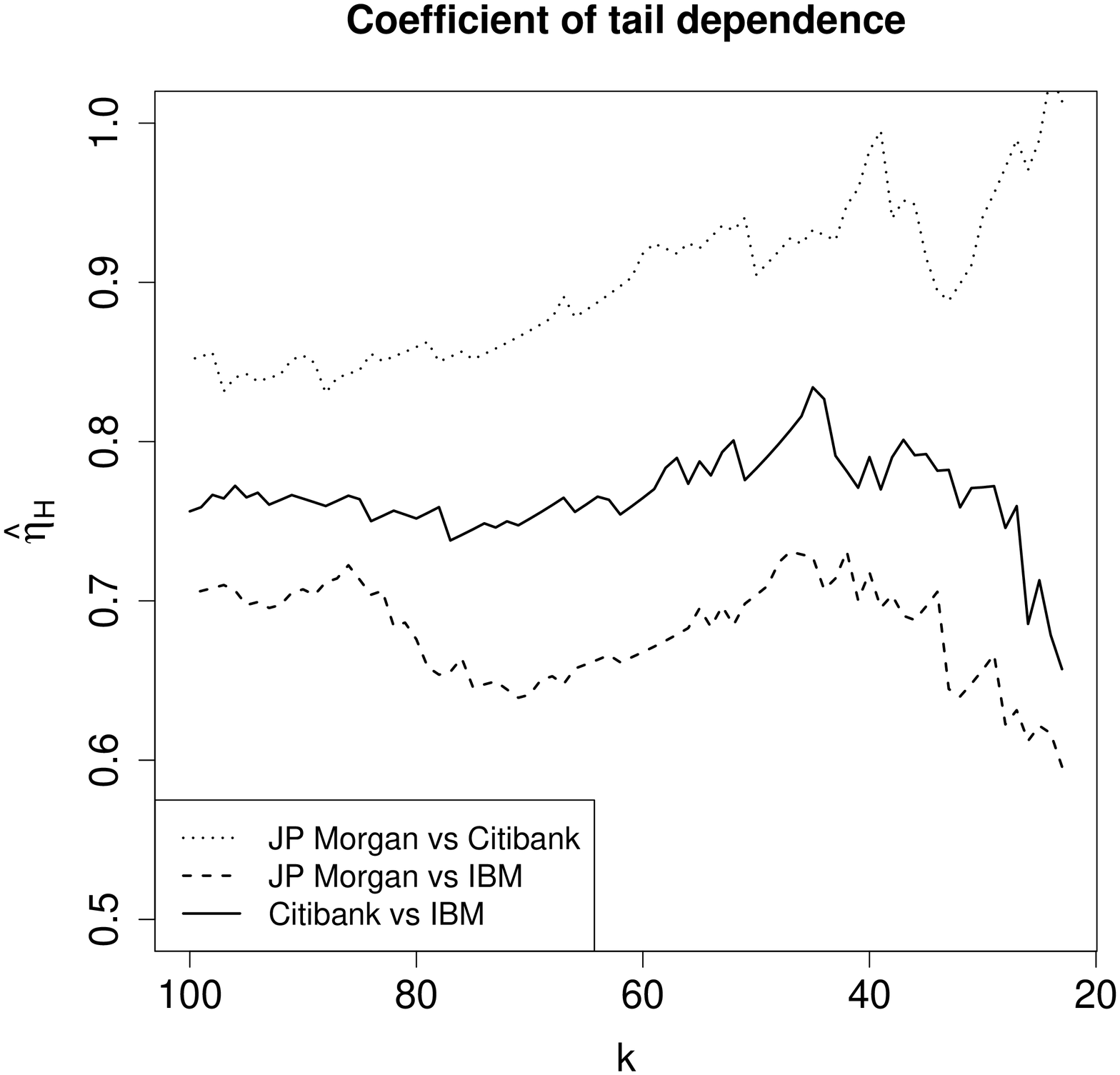}}
\caption[Estimators of the tail dependence coefficient $\widehat{\chi} (u)$ and of the coefficient of tail dependence $\widehat{\eta}_H$]{Estimators $\widehat{\chi} (u)$ (left) and $\widehat{\eta}_H$ (right) for the three pairs of log-returns.}
\label{fig:chiplot}
\end{figure}

Another frequently used measure in case of asymptotic independence can be obtained from \eqref{eq:eta2}.
Taking logarithms on both sides and using slow variation of $\mathcal{L}$, we obtain
\begin{equation*}
\frac{1}{\eta} 
= \lim_{u \uparrow 1} 
\frac{\log{\Pr[F_X (X) > u, \, F_Y (Y) > u]}}{\log \Pr[F_X (X) > u]}.
\end{equation*}
\citet{coles1999} proposed the coefficient
\begin{equation*}
\overline{\chi} 
\coloneqq \lim_{u \uparrow 1} \overline{\chi} (u) 
= \lim_{u \uparrow 1} \frac{2 \log{ \PP [F_X (X) > u]}}{\log \PP [F_X (X) >u, F_Y (y)>u ]} - 1 = 2 \eta - 1. 
\end{equation*}
As $0 < \eta \le 1$, we have $-1 < \overline{\chi} \leq 1$. For a bivariate normal dependence structure, $\overline{\chi}$ is equal to the correlation coefficient $\rho$.
Together, the two coefficients $\chi$ and $\overline{\chi}$ contain complementary information on the amount of tail dependence:
\begin{itemize}
\item
If $\chi = 0$ and $-1 < \overline{\chi} \leq 1$, the variables $X$ and $Y$ are asymptotically independent and $\overline{\chi}$ quantifies the degree of dependence. The cases $\overline{\chi} >0$, $\overline{\chi} = 0$, and $\overline{\chi} <0$ correspond to positive association, near independence, and negative association, respectively.
\item
If $\overline{\chi} = 1$ and $0 < \chi \leq 1$, the variables are asymptotically dependent, and $\chi$ quantifies the degree of dependence. 
\end{itemize}

Although the graphical procedures in this section are easy to use, it is difficult to draw a conclusion from the plots, and it might be better to consider formal tests of asymptotic independence (Section~\ref{ss:ai:test}).
For an overview of more recent and advanced techniques in asymptotic independence, see for example \citet{carvalho2012} or \citet{bacro2013}. 

\subsection{Testing for asymptotic (in)dependence}
\label{ss:ai:test}

Above we have called the variables $X$ and $Y$ asymptotically independent if $\ell(x, y) = x + y$. A sufficient condition for asymptotic independence is that the coefficient of tail dependence, $\eta$, in \eqref{eq:eta2} is less than unity. 

These observations yield two possible approaches to test for asymptotic (in)dependence, depending on the choice of the null hypothesis. Suppose the function $\mathcal{L}$ in \eqref{eq:eta2} converges to a positive constant.
\begin{itemize}
\item
Testing $H_0 : \eta = 1$ versus $H_1 : \eta < 1$ amounts to choosing asymptotic dependence as the null hypothesis.
\item
Testing $H_0 : \forall (x, y), \; \ell(x, y) = x + y $ versus $H_1 : \exists (x, y), \; \ell(x, y) < x + y$ means that asymptotic independence is taken as the null hypothesis.
\end{itemize}

\paragraph{Asymptotic dependence as the null hypothesis.}
\index{asymptotic independence!testing for asymptotic independence}

In order to test for $H_0 : \eta = 1$ versus $H_1 : \eta < 1$, the idea is to estimate $\eta$ and to reject $H_0$ if the estimated value of $\eta$ is below some critical value. The estimator chosen in \citet{draisma2004} is the maximum likelihood estimator, $\widehat{\eta}_{\mathrm{MLE}}$, obtained by fitting a generalized Pareto distribution to the $k$ highest order statistics of the sample $\widehat{T}_1,\ldots,\widehat{T}_n$ in \eqref{eq:Tmin}. The effective sample size, $k$, tends to infinity at a slower rate than $n$.

In order to derive the asymptotic distribution of $\widehat{\eta}_{\mathrm{MLE}}$, condition~\eqref{eq:eta2} needs to be refined. Consider the function
\[
  q(x, y) \coloneqq \Pr[1 - F_X (X) < x, 1 - F_Y (Y) < y].
\]
Note that the tail dependence coefficient $\chi$ in \eqref{eq:chi} is given by $\chi = \lim_{t \downarrow 0} q(t, t) / t$. Equation~\eqref{eq:eta2} implies that the function $t \mapsto q(t, t)$ is regularly varying at zero with index $1/\eta$, i.e., $\lim_{t \downarrow 0} q(tx, tx) / q(t, t) = x^{1/\eta}$ for $x > 0$. In \citet{draisma2004}, this relation is refined to \emph{bivariate regular variation}: it is assumed that
\begin{equation}
\label{eq:eta:biv}
  \lim_{t \downarrow 0} \frac{q (tx,ty)}{q(t,t)} = c(x,y), 
  \qquad (x, y) \in (0, \infty)^2.
\end{equation} 
The limit function $c$ is homogeneous of order $1/\eta$, i.e., $c(ax, ay) = a^{1/\eta} c(x, y)$ for $a > 0$. Clearly, $c(1, 1) = 1$. To control the bias of $\widehat{\eta}_{\mathrm{MLE}}$, a second-order refinement of \eqref{eq:eta:biv} is needed: assume the existence of the limit
\begin{equation*}
  c_1 (x,y) 
  \coloneqq \lim_{t \downarrow 0} \frac{\frac{q(tx,ty)}{q (t,t)} - c(x,y)}{q_1 (t)},
\end{equation*}
for all $x,y \geq 0$ with $x + y > 0$. Here, $0 < q_1(t) \to t$ as $t \to 0$, while the limit function $c_1$ is neither constant nor a multiple of $c$. The convergence is assumed to be uniform on $\{(x,y) \in [0,\infty)^2 : x^2 + y^2 = 1 \}$.

Suppose that the function $c$ has partial derivatives $c_x \coloneqq \partial c(x,y) / \partial x$ and $c_y \coloneqq \partial c(x,y) / \partial y$. 
Let $k$ be a sequence such that $\sqrt{k} \, q_1 (q^{-1}(k/n,k/n)) \to 0$ as $n \to \infty$, where $q^{-1}(u, u) = t$ if and only if $q(t, t) = u$. 
Then $\widehat{\eta}_{\mathrm{MLE}}$ is asymptotically normal,
\begin{equation*}
  \sqrt{k} (\widehat{\eta}_{\mathrm{MLE}} - \eta) \dto \mathcal{N} \bigl(0, \sigma^2(\eta)\bigr),
\end{equation*}
the asymptotic variance being
\[
  \sigma^2(\eta) 
  = (1 + \eta)^2 \, (1 - \chi) \, \bigl(1 - 2 \chi c_x (1,1) c_y (1,1) \bigr).
\]
Under the null hypothesis $H_0 : \eta = 1$, consistent estimators of $\chi$ and $c_x (1,1)$ are given by
\begin{align*}
  \widehat{\chi} 
  &\coloneqq \frac{k \widehat{T}_{(k+1)}}{n}, &
  \widehat{c}_x (1,1) 
  &\coloneqq \frac{\widehat{p}^{\, 5/4}}{n} 
  \left[ \widehat{T}_{(k+1)}^{(n,\widehat{p}^{\, -1/4})} - \widehat{T}_{(k+1)} \right],
\end{align*}
where $\widehat{p} \coloneqq k/ \widehat{\chi}$ and $\widehat{T}_{(k+1)}^{(n,u)}$ is the $(k+1)$th largest observation of 
\begin{equation*}
\widehat{T}_{i}^{(n,u)} \coloneqq \min \left( \frac{ (1 + u)}{1 - \widehat{F}_X (X_i)}, \frac{1}{1 - \widehat{F}_Y (Y_i)} \right), \qquad i = 1,\ldots,n.
\end{equation*}
The estimator $\widehat{c}_y (1,1)$ is defined analogously to $\widehat{c}_x (1,1)$. The null hypothesis $H_0 : \eta = 1$ is rejected at significance level $\alpha$ if
\begin{equation}
\label{eq:etatest}
  \widehat{\eta}_{\mathrm{MLE}} \le 1 - \frac{\widehat{\sigma}}{\sqrt{k}} \Phi^{-1}(1-\alpha),
\end{equation}
where $\widehat{\sigma} = \widehat{\sigma}(1)$ or $\widehat{\sigma} = \widehat{\sigma}(\hat{\eta}_{\mathrm{MLE}})$. 

Figure~\ref{fig:hyptest1} shows the estimates of $\widehat{\eta}_{\mathrm{MLE}}$ for a varying number of order statistics $k$, together with the lines defining the critical regions for the two tests in \eqref{eq:etatest}. We used the function \textsf{gpd.fit} from the \textsf{ismev} package \citep{ismev}. We see clearly that for JP Morgan versus Citibank, for every threshold, asymptotic dependence cannot be rejected; for JP Morgan versus IBM, asymptotic dependence is rejected at every threshold; and for Citibank versus IBM, results vary depending on the value of $k$.

\begin{figure}[ht]
\centering
\subfigure{\includegraphics[width=0.32\textwidth]{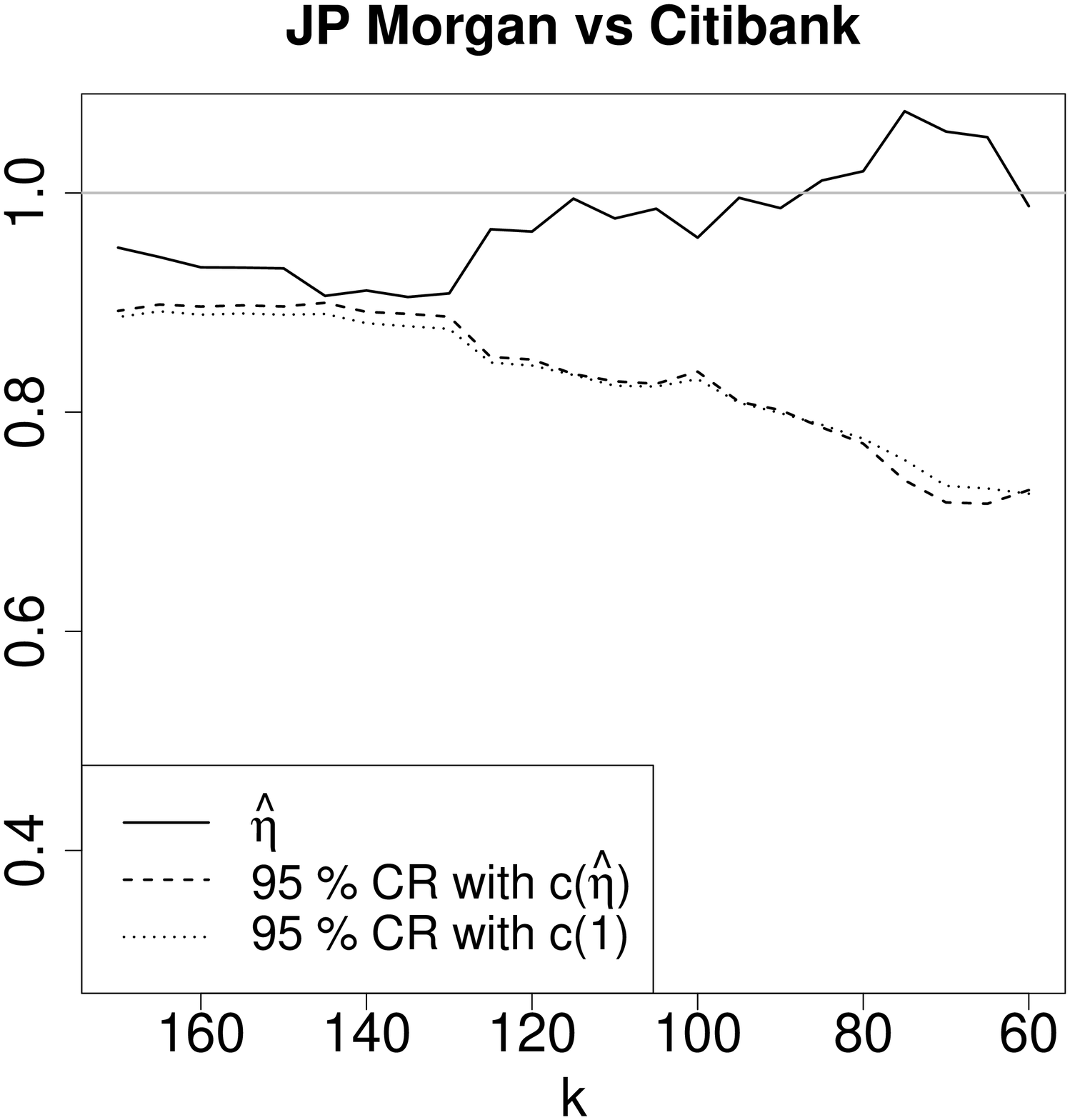}}
\subfigure{\includegraphics[width=0.32\textwidth]{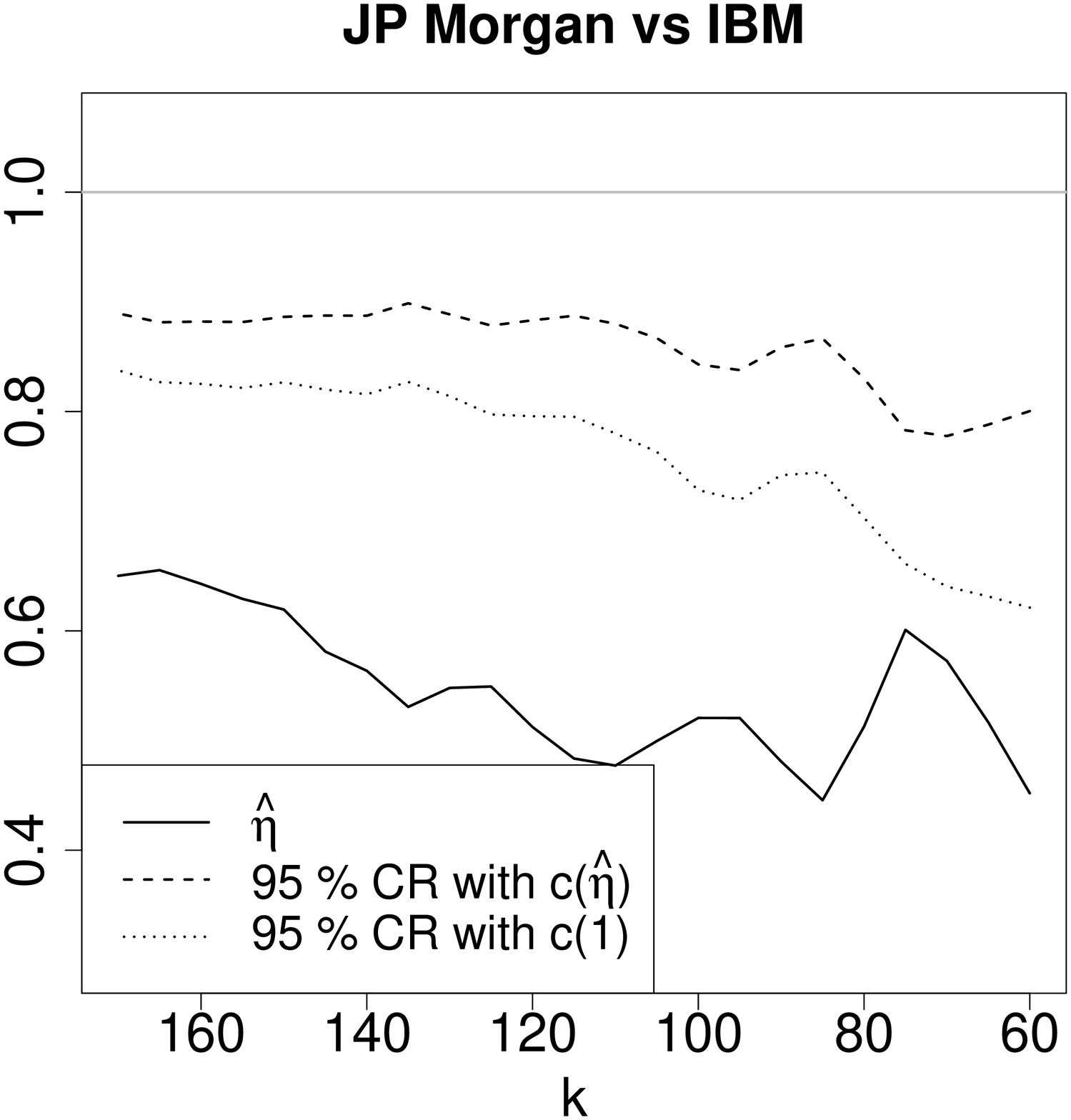}}    
\subfigure{\includegraphics[width=0.32\textwidth]{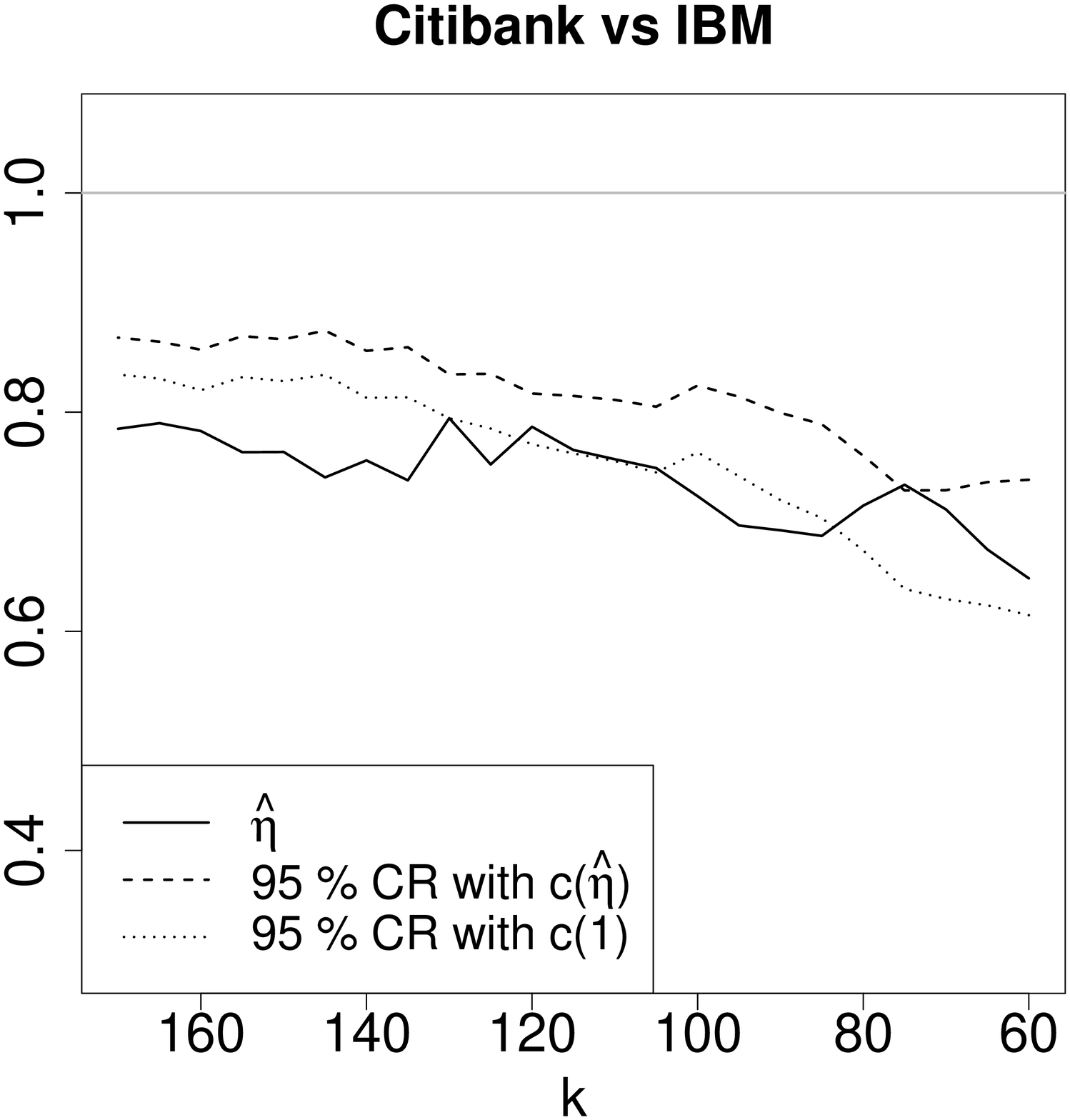}}
\caption[Estimates $\widehat{\eta}_{MLE}$ of the coefficient of tail dependence with $95 \%$ critical regions for a null hypothesis of asymptotic dependence]{Estimates $\widehat{\eta}$ (solid lines) obtained by maximum likelihood based on excesses over the threshold $\widehat{T}_{(k+1)}$, together with pointwise $95 \%$ critical regions (CR).}
\label{fig:hyptest1}
\end{figure}

\paragraph{Asymptotic independence as the null hypothesis.}
\index{asymptotic independence!testing for asymptotic independence}

Another approach for deciding between asymptotic dependence and asymptotic independence is to assume a null hypothesis of asymptotic independence, i.e., $H_0: \forall (x, y), \; \ell(x,y) = x + y$. Recall the nonparametric estimator $\widehat{\ell}$ in Section~\ref{ss:estim:stdf}. A natural approach would be to reject $H_0$ as soon as the difference between  $\widehat{\ell}$ and the function $(x, y) \mapsto x + y$ is too large. However, from \citet{einmahl2006} it follows that, under $H_0$,
\begin{equation*}
  \sup_{x,y \in [0,1]} 
  \sqrt{k} \abs{ \widehat{\ell} (x,y) - (x + y) } \pto 0.
\end{equation*}
The limit being degenerate, it is not possible to compute critical values.

In \citet{huesler2009} another estimator of $\ell$ is proposed, based on a division of the sample into two sub-samples. Note first that we can write $\widehat{\ell}$ as 
\begin{equation*}
  \widehat{\ell} (x,y) 
  = \frac{1}{k} \sum_{i = 1}^n 
  \1 \left\{ 
    X_i > \widehat{F}_X^{-1} \left( 1 - \frac{kx}{n} \right) \text{ or } 
    Y_i > \widehat{F}_Y^{-1} \left( 1 - \frac{ky}{n} \right) 
  \right\}.
\end{equation*}
For convenience, assume that $n$ is even. The first sub-sample $X_1,\ldots,X_{n/2}$ is compared with the $[kx]$-th largest order statistics of the second sub-sample $X_{n/2 + 1},\ldots,X_n$; in other words, we use the estimator
\begin{align*}
\widetilde{\ell}_n (x,y) & = \frac{1}{k} \sum_{i = 1}^{n/2} \mathbbm{1} \left\{ X_i > \widehat{\widetilde{F}}_X^{-1} \left( 1 - \frac{2kx}{n} \right) \text{ or } Y_i > \widehat{\widetilde{F}}_Y^{-1} \left( 1 - \frac{2ky}{n} \right) \right\} \\
& = \frac{1}{k} \sum_{i = 1}^{n/2} \mathbbm{1} \left\{ \widetilde{R}_{i,X} > n/2 + 1 - kx \text{ or } \widetilde{R}_{i,Y} > n/2 + 1 - ky \right\},
\end{align*} 
where $\widetilde{R}_{i,X}$ for $i=1,\ldots,n/2$ denotes the rank of $X_i$ among $X_{n/2+1},\ldots,X_n$. 
Under $H_0$, the estimator $\widetilde{\ell}_n (x,y)$ converges in probability to $x + y$, for fixed $x,y > 0$. Define
\begin{equation*}
D_n (x,y) \coloneqq \sqrt{k} \{ \widetilde{L}_n (x,y) - (x + y) \}, \qquad x,y \in [0,1].
\end{equation*}
Then under $H_0$, assuming certain regularity conditions on $\ell$ and $k$ \citep[page 992]{huesler2009}, if $k= k_n \to \infty$ as $n \to \infty$,
\begin{equation*}
\left\{ D_n (x,y) \right\}_{x,y \in [0,1]} \dto 
\left\{ W_1 (2x) + W_2 (2y) \right\}_{x,y \in [0,1]}, \qquad \text{as } n \to \infty,
\end{equation*}
where $W_1$ and $W_2$ are two independent Brownian motions. By the continuous mapping theorem,
\begin{align*}
  T_{I,n} 
    & \coloneqq \int_{[0,1]^2} D_n (x,y)^2 \, \diff x \, \diff y 
    \dto \int_{[0,1]^2} \bigl( W_1 (2x) + W_2 (2y) \bigr)^2 \, \diff x \, \diff y
    \eqqcolon T_I, \\
  T_{S,n} 
    & \coloneqq \sup_{x,y \in [0,1]} \abs{ D_n (x,y) } 
    \dto \sup_{x,y \in [0,1]} \abs{ W_1 (2x) + W_2 (2y) } 
    \eqqcolon T_S.
\end{align*}
We reject the null hypothesis of asymptotic independence at significance level $\alpha$ if $T_{I,n} > Q_{T_I}(1-\alpha)$ or if $T_{S,n} > Q_{T_S}(1-\alpha)$, where $Q_{T_I}$ and $Q_{T_S}$ represent the quantile functions of $T_I$ and $T_S$ respectively. \citet{huesler2009} compute $Q_{T_I}(0.95) = 6.237$ and $Q_{T_S}(0.95) = 4.956$. 

Figure~\ref{fig:hyptest2} shows the values of the test statistics $T_{I,n}$ and $T_{S,n}$ for the three pairs of stock returns, together with the pointwise critical regions for a range of $k$-values. The results are in agreement with Figure~\ref{fig:hyptest1}: asymptotic independence is rejected for JP Morgan versus Citibank, asymptotic independence cannot be rejected for JP Morgan versus IBM, and conclusions for Citibank versus IBM depend on the value of $k$.

\begin{figure}[ht]
\centering
\subfigure{\includegraphics[width=0.48\textwidth]{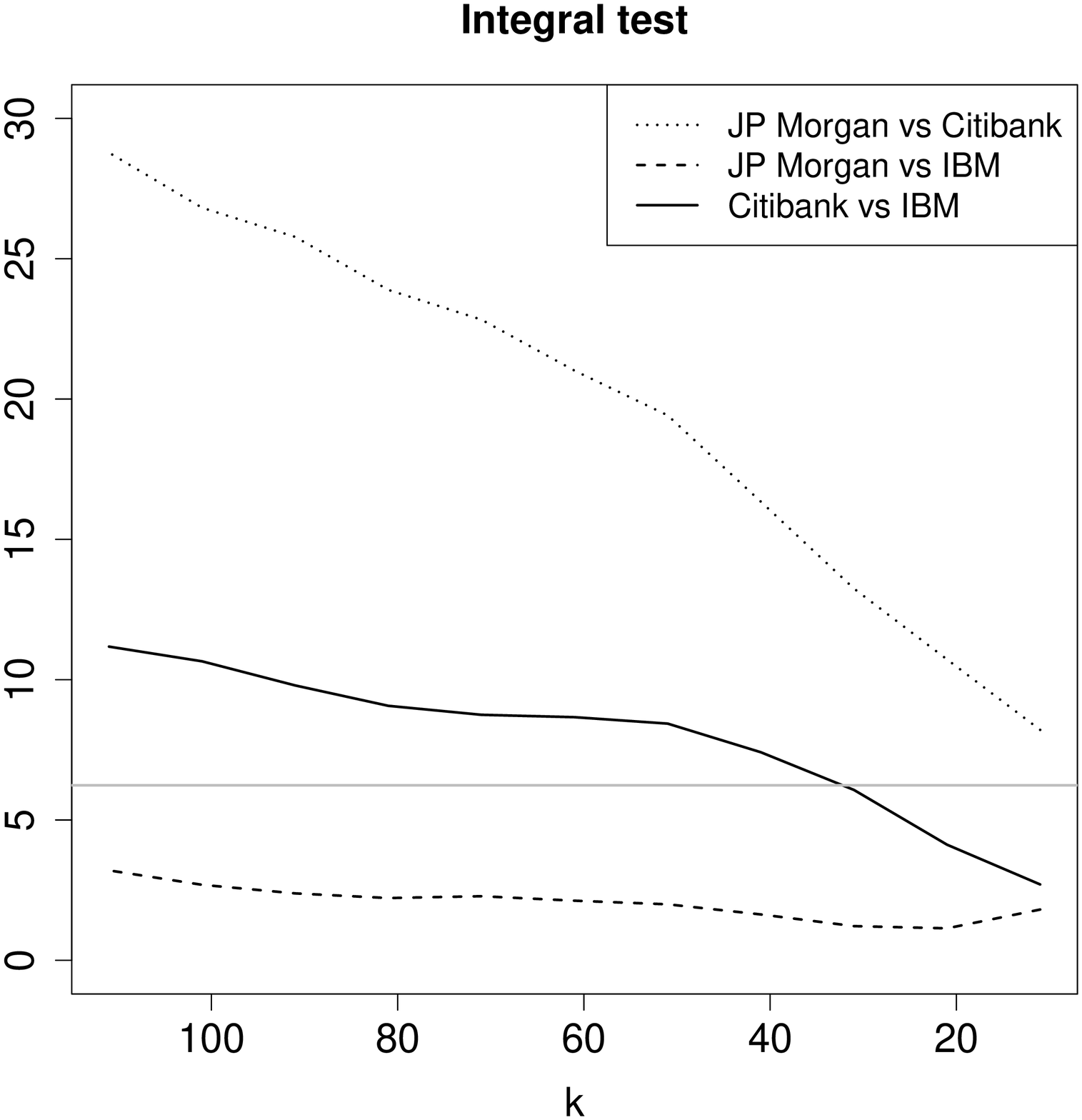}}
\subfigure{\includegraphics[width=0.48\textwidth]{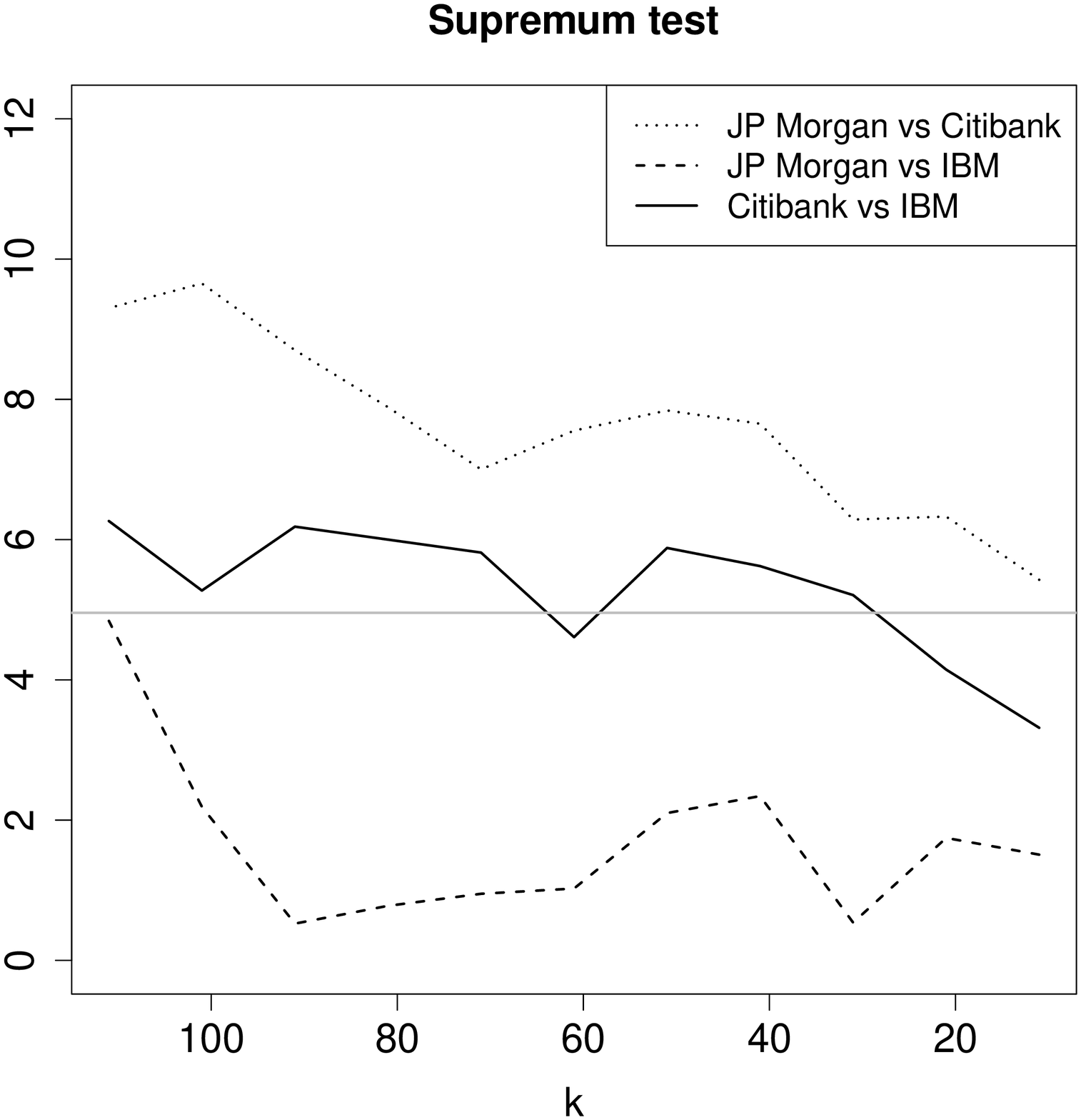}}    
\caption[Test statistics for a null hypothesis of asymptotic independence]{Test statistics $T_{I,n}$ (left) and $T_{S,n}$ (right) for a varying number of order statistics $k$. The horizontal lines define the critical regions $Q_{T_I}(0.95) = 6.237$ and $Q_{T_S}(0.95) = 4.956$.}
\label{fig:hyptest2}
\end{figure}

\section*{Acknowledgments}

The authors gratefully acknowledge funding by contract ``Projet d'Act\-ions de Re\-cher\-che Concert\'ees'' No.\ 12/17-045 of the ``Communaut\'e fran\c{c}aise de Belgique'' and by IAP research network Grant P7/06 of the Belgian government (Belgian Science Policy). Anna Kiriliouk and Micha{\l} Warcho{\l} gratefully acknowledge funding from the Belgian Fund for Scientific Research (F.R.S. - FNRS).

\bibliographystyle{chicago}
\renewcommand\refname{REFERENCES} 
\bibliography{NonParEst}

\end{document}